\documentstyle[aps]{revtex}
\begin{document}
%\documentstyle[12pt]{article} %\documentstyle[psfig,prd,aps]{revtex}
%\input epsf
%\tightenlines
 \def\BA{\Bbb A}
 \def\BC{\Bbb C}
 \def\BR{\Bbb R}
 \newfont{\titlefont}{cmssbx10 scaled\magstep5}
%%-------------------------------------
\newcommand{\CA}{{\cal A}}
\newcommand{\CB}{{\cal B}}
\newcommand{\CC}{{\cal C}}
\newcommand{\CD}{{\cal D}}
\newcommand{\CE}{{\cal E}}
\newcommand{\CF}{{\cal F}}
\newcommand{\CG}{{\cal G}}
\newcommand{\CH}{{\cal H}}
\newcommand{\CI}{{\cal I}}
\newcommand{\CJ}{{\cal J}}
\newcommand{\CK}{{\cal K}}
\newcommand{\CL}{{\cal L}}
\newcommand{\CM}{{\cal M}}
\newcommand{\CN}{{\cal N}}
\newcommand{\CO}{{\cal O}}
\newcommand{\CP}{{\cal P}}
\newcommand{\CQ}{{\cal Q}}
\newcommand{\CR}{{\cal R}}
\newcommand{\CS}{{\cal S}}
\newcommand{\CT}{{\cal T}}
\newcommand{\CU}{{\cal U}}
\newcommand{\CV}{{\cal V}}
\newcommand{\CW}{{\cal W}}
\newcommand{\CX}{{\cal X}}
\newcommand{\CY}{{\cal Y}}
\newcommand{\CZ}{{\cal Z}}
%%-------------------------- %%  Command Abbreviations
%%--------------------------
\newcommand{\bea}{\begin{eqnarray}} \newcommand{\eea}{\end{eqnarray}}
\newcommand{\beqa}{\begin{eqnarray}} \newcommand{\eeqa}{\end{eqnarray}}
\newcommand{\beq}{\begin{equation}} \newcommand{\eeq}{\end{equation}}
\newcommand{\non}{\nonumber} \newcommand{\eqn}[1]{\beq {#1}\eeq}
\newcommand{\eu}{Euclidean\ }
\newcommand{\lmk}{\left(} \newcommand{\rmk}{\right)}
\newcommand{\lkk}{\left[} \newcommand{\rkk}{\right]}
\newcommand{\lhk}{\left \{ } \newcommand{\rhk}{\right \} }
\newcommand{\lnk}{\left \{ } \newcommand{\rnk}{\right \} }
\newcommand{\del}{\partial} \newcommand{\abs}[1]{\vert{#1}\vert}
\newcommand{\vect}[1]{\mbox{\boldmath${#1}$}}
\newcommand{\bib}{\bibitem} \newcommand{\new}{\newblock}
\newcommand{\la}{\left\langle} \newcommand{\ra}{\right\rangle}
\newcommand{\bfx}{{\bf x}} \newcommand{\bfk}{{\bf k}}
\newcommand{\vex}{{\vect x}} \newcommand{\vek}{{\vect k}}
\newcommand{\vep}{{\vect p}} \newcommand{\veq}{{\vect q}}
\newcommand{\veg}{{\vect \gamma}}
\newcommand{\gtilde}{~\mbox{\raisebox{-1.0ex}{$\stackrel{\textstyle >}
{\textstyle \sim}$ }}}
\newcommand{\ltilde}{~\mbox{\raisebox{-1.0ex}{$\stackrel{\textstyle <}
{\textstyle \sim}$ }}}
\newcommand{\gsim}{~\mbox{\raisebox{-1.0ex}{$\stackrel{\textstyle >}
{\textstyle \sim}$ }}}
\newcommand{\lsim}{~\mbox{\raisebox{-1.0ex}{$\stackrel{\textstyle <}
{\textstyle \sim}$ }}}
\newcommand{\mh}{m_H} \newcommand{\veff}{V_{\rm eff}}
\newcommand{\phip}{\phi_+}
\newcommand{\eff}{\rm eff}
\newcommand{\Ghat}{\hat{\Gamma}}
\newcommand{\mpl}{M_{Pl}}
\newcommand{\gn}{g_{*N}}

\draft
%@twocolumnfalse\endcsname
\preprint{SU-ITP-98/52, YITP-98-49, hep-ph/9809409}
\title{Is Warm Inflation Possible?}
\author{Jun'ichi Yokoyama}
\address{Department of Physics, Stanford University, Stanford, CA
94305-4060 and\\ Yukawa Institute for Theoretical Physics, Kyoto
University, Kyoto 606-8502, Japan}
\author{Andrei Linde}
\address{Department of Physics, Stanford University, Stanford, CA
94305-4060, USA}
\date{August 17, 1998}
\maketitle
\begin{abstract}
We show that it is extremely difficult and perhaps even impossible to
have inflation supported by thermal effects.
\end{abstract}
\pacs{PACS: 98.80.Cq  \hskip 2.2cm SU-ITP-98-52 \hskip 1cm YITP-98-49
\hskip 2.2cm  hep-ph/9809409}
%\vskip2pc]

%%%%%%%%%%%%%%%%%%%%%%%%%%%%%%%%%%%%%%%%%%%%%%%%%%%%%%%%%%%%%%%%%%%%%%%%%
\section{ Introduction} \label{intr}
%%%%%%%%%%%%%%%%%%%%%%%%%%%%%%%%%%%%%%%%%%%%%%%%%%%%%%%%%%%%%%%%%%%%%%%%

First models of inflationary cosmology \cite{Guth,newinf} were based
on an assumption that inflation appears as a result of
high-temperature phase transitions with supercooling in the early
universe \cite{Kirzhnits}. This idea survived  less than two years,
after which it was replaced by chaotic inflation   which does not
require initial thermal equilibrium and phase transitions
\cite{chaoinf}. There were several reasons why this happened; here is
a short list \cite{book}:

\begin{itemize}
\item The assumption that the universe was hot from the very
beginning is not necessary in the chaotic inflation scenario.
Moreover, the assumption that inflation begins only after a long
stage of cooling implies that the universe from the very beginning
must be very large and homogeneous. This means that such models do
not provide a complete solution to the homogeneity and flatness
problems.
\item The theories where inflation occurs after supercooling require fine
tuned effective potentials satisfying ``thermal constraints''.
\item The inflaton field typically has a very weak interaction with other
fields, so it may be out of thermal equilibrium in the early
universe.
\item Even if the inflaton field is in thermal equilibrium, it takes
a lot of time for it to roll to the minimum of the
temperature-dependent effective potential. In many cases, inflation
occurs while the field rolls down, so when it arrives to the minimum
of the effective potential at $\phi = 0$, the temperature vanishes.
\item Even if the temperature does not completely vanish when the
field falls
to the
minimum of the effective potential, it  vanishes during supercooling,
so the thermal effects become irrelevant for the description of the
tunneling and of the subsequent stage of slow rolling of the inflaton
field \cite{Therm}.
\end{itemize}
For all of these reasons, the idea that inflation is somehow related
to high-temperature effects was almost completely abandoned. However,
recently there was an attempt to revive it in the context of the
so-called  warm inflation scenario \cite{warm}.

The main idea of the scenario can be formulated as follows. Suppose
the universe is in a state of thermal equilibrium, and the field
$\phi$ slowly rolls down to its minimum. When the universe expands,
its temperature tends to decrease as $a(t)^{-1}$, where $a$ is the
scale factor. Therefore one   expects that the temperature in an
inflationary universe falls down exponentially and immediately
becomes irrelevant. However, if the scalar field interacts with other
particles, it may transfer some of its energy to the thermal bath.
This may keep the temperature from falling to zero.

This regime may continue in a self-consistent way if the  amount of
particles produced due to the interaction of the scalar field $\phi$
with thermal bath is large enough to keep the field from a rapid fall
to the minimum of the effective potential.

However, we immediately see a lot of problems here.   During each
time $H^{-1}$ the universe expands $e$ times, and the energy density
of previously existed ultrarelativistic particles becomes $e^{-4}$
times smaller. Therefore most of the particles during warm inflation
at any given moment should have been created during the previous time
interval   $\sim 0.2 H^{-1}$.   If the energy release is going to
keep the field from rapidly falling down, then the energy released
each  time $0.2 H^{-1}$  must be very small. Otherwise the field
rapidly falls down, and there is no inflation.  But if the energy
release is small, then the total number of particles in the warm
universe must be very small, and therefore their interaction with the
scalar field $\phi$ may be too small to keep it from rapid falling
down.

Despite this problem, the basic idea is rather interesting and it
should not be discarded without a serious investigation. It would be
very interesting to see, in particular, whether this idea could help
to realize inflationary universe scenario in the context of string
theory.

 The results of investigation  of warm inflation indicated, as
expected, that this regime  is very hard to obtain \cite{bgr}.
However,  the methods used in this investigation were rather complex
and not very intuitive.   Therefore the basic mechanism of warm
inflation remained obscure, and there remained a hope that one can
obtain a good realization of this scenario by considering slightly
more complicated theories of elementary particles.

In this paper we will try to give a simple and intuitive explanation
of the mechanism of dissipation of the energy of the inflaton field,
which will help us to understand the main problems of the warm
inflation scenario. Then we will confirm our expectations using more
rigorous methods of non-equilibrium quantum statistics.

\section{ Intuitive argument} \label{intuit}
In order to explain the basic idea of warm inflation, as well as its
shortcomings, in this section we will give a simple and intuitive
discussion of this scenario in the theory of a massive
  inflaton field $\phi$ interacting with a massless field $\chi$:
\begin{equation}\label{1}
L =  {1\over 2} (\partial_\mu \phi)^2  + {1\over 2} (\partial_\mu
\chi)^2 - {m^2\over 2}   \phi^2 -{g^2\over 2}   \phi^2 \chi^2 \ .
\end{equation}
If one neglects interaction between these two fields, equation of
motion for the scalar field $\phi$ has a familiar form
\begin{equation}\label{2a}
\ddot \phi   + 3 H \dot\phi = -m^2\phi .
\end{equation}
The  term $3 H \dot\phi$ represents energy loss of the   field $\phi$
due to expansion of the universe. This term, which is similar to the
viscosity term in an equation of motion of a pendulum in a viscous
liquid,  slows down the motion of the field $\phi$. As a result, the
potential energy of the field changes very slowly, which under some
circumstances may lead to inflation \cite{book}.

  One could expect that interaction of the fields $\phi$ and $\chi$
may lead to an additional energy loss of the field $\phi$ due to
creation of $\chi$-particles (reheating), which can be represented by
adding another friction term to this equation \cite{ASTW,MoriSasa}:
\begin{equation}\label{2}
\ddot \phi   +\Gamma \dot\phi+ 3 H \dot\phi = -m^2\phi .
\end{equation}
If $\Gamma \gg H$, the field $\phi$ will roll down   more slowly than
in the absence of interaction. Therefore one could expect that
inflation may continue for a longer time \cite{ASTW}, and may occur
even in such theories where otherwise it would be impossible
\cite{YM}.

Further development of the theory of reheating has shown that the
situation is more complicated. Addition of the term $\Gamma \dot\phi$
 effectively describes energy
loss due to particle creation only at the stage of oscillations of
the scalar field $\phi$  \cite{reheat}, and only in the case when
these oscillations do not lead to parametric resonance \cite{KLS}. It
would be inappropriate to use this equation for the description of
energy loss of the field $\phi$ during inflation, assuming, as usual,
that density of all particles during inflation is exponentially
small.

However, if the universe is hot, and there are many $\chi$-particles
in the thermal bath, then in a certain approximation the motion of
the scalar field $\phi$ is indeed described by Eq. (\ref{2}). This
effect is the basic ingredient of the warm inflation scenario. The
standard description of this effect is very complicated because it
involves summation of series of higher order diagrams in the
non-equilibrium quantum statistics \cite{Mor,GR}. We will describe
this method later. In this section we would like to give a simple
interpretation of this effect which will simultaneously tell us
whether this effect can be  significant enough to give rise to an
inflationary regime.

Let us write equation of motion for the field $\phi$ taking into
account interaction between the scalar fields $\phi$ and $\chi$ in
the Hartree approximation:
\begin{equation}\label{PPP}
 \ddot{\phi}+3H\dot{\phi}+m^2\phi+g^2\phi\langle\chi^2\rangle=0.
\end{equation}
In a state of thermal equilibrium one has \cite{Kirzhnits,book}
\begin{equation}
 \langle\chi^2\rangle_{eq}=
{1\over {(2\pi)^3}}\int \frac{n_{\chi}^{eq}(p) d^3p} {\omega_\chi} \
,
\end{equation}
where
\begin{equation}
n_{\chi}^{eq}=  \frac{1}{\exp{\omega_\chi\over T}-1} \ , ~~~~~
\omega_\chi = \sqrt{p^2 + g^2\phi^2} \ .
\end{equation}

We will assume that $g\phi \ll T$, because in the opposite limit all
thermal effects are exponentially small. In this case
\begin{equation}\label{1b}
 \langle \chi^2\rangle_{eq} = {T^2\over 12}\left(1  - {3g\phi\over \pi
T}\right) +...
\end{equation}

Now suppose the scalar field $\phi$ changes by $\Delta \phi$. This
changes masses of $\chi$-particles $g\phi$, and the equilibrium value
of $n_{\chi}^{eq}$ should change correspondingly.  However, this
change cannot happen instantaneously; it occurs with a   time delay
$\Delta t \sim \Gamma_\chi^{-1}$, where  $\Gamma_\chi$ is a decay
width of $\chi$-particles at finite temperature. This means that
$n_{\chi} $ deviates from its equilibrium value $n_{\chi}^{eq}$: when
the scalar field reaches its value $\phi$, the occupation number
remains equal to $n_{\chi}^{eq}$ at earlier time, when the field was
equal to $\phi - \Delta \phi \sim \phi - \dot\phi \Delta t$, with
$\Delta t \sim \Gamma_\chi^{-1}$ \cite{HS}:
\begin{equation}
\Delta n_{\chi} = n_{\chi} - n_{\chi}^{eq} \sim -{dn_{\chi}^{eq}\over
d\phi}\, \dot\phi\,  \Gamma_\chi^{-1} \ .
\end{equation}
Therefore   $\langle\chi^2\rangle$ in equation (\ref{PPP}) slightly
differs from  $\langle\chi^2\rangle_{eq}$:
\begin{equation}
\langle\chi^2\rangle= {1\over {(2\pi)^3}}\int \frac{n_{\chi}(p) d^3p}
{\omega_\chi} =  \langle\chi^2\rangle_{eq} - {\dot\phi\over
{(2\pi)^3}}\int \frac{d^3p}{\omega_\chi}\, {{dn_{\chi}^{eq}\over
d\phi}\, \Gamma_\chi^{-1}}\ .
\end{equation}

Note that $\Gamma_\chi^{-1}$ depends on   momentum $p$,   it is small
for $p \ll T$.  The leading contribution to the last integral is
given by particles with typical momentum $p \sim T$. In this case, as
we will show in Sect. \ref{diss},
\begin{equation}\label{??}
 \Gamma_\chi \simeq \frac{g^4 T}{192\pi} \ .
\end{equation}
Evaluation of  the last integral yields:
\begin{equation}
\langle\chi^2\rangle= \langle\chi^2\rangle_{eq} + {C \phi\over g^2T
}\dot\phi = {T^2\over 12}\left(1  - {3g\phi\over \pi T}\right) + {C
\phi\over g^2T }\dot{\phi}+ ...\ .
\end{equation}
where $C = O(10)$.

With an account taken of this correction, Eq. (\ref{PPP}) can be
represented in the following form \cite{HS}:
\begin{equation}\label{PPPP}
 \ddot{\phi}+{C  \phi^2\over  T } \dot\phi +
3H\dot{\phi}+m^2\phi+g^2\phi\langle\chi^2\rangle_{eq}~ \approx ~
\ddot{\phi}+{C \phi^2\over  T } \dot\phi + 3H\dot{\phi}+m^2\phi+{g^2
T^2\over 12}\phi  ~ = ~0.
\end{equation}
In Sect. \ref{diss} we will obtain this equation by a more rigorous,
but somewhat more complicated method. The simple and intuitive
approach which we have used in this section allows us to easily
understand the nature of the viscosity term ${C \phi^2\over T }
\dot\phi$ and evaluate its possible significance.

Indeed, at the first glance this term could be very large. In fact,
it is not suppressed by any powers of the small coupling constant
$g$, and in certain cases ${C  \phi^2\over  T }\dot\phi$ may be
greater than the usual viscosity term $3H\dot\phi$. This was the main
reason why it was expected that this additional viscosity term could
play an important role in inflationary theory.

However, let us  remember the procedure of the derivation of the term
${C \phi^2\over  T } \dot\phi$. The leading term in the expression
for $\langle\chi^2\rangle_{eq}$ in Eq. (\ref{1b}) is
$\phi$-independent: $\langle\chi^2\rangle_{eq} \approx T^2/12$. The
term ${C \phi\over g^2 T }\dot\phi$ appears because the {\it
subleading} term in Eq. (\ref{1b})  depends on $\phi$, and because
this term {\it slightly} differs from its thermally equilibrium value
${g\phi T \over 4\pi}$. Thus the term ${C \phi^2\over  T } \dot\phi$
appears in Eq. (\ref{PPPP}) as a correction to a correction, because
${C \phi\over g^2T }\dot\phi  \ll {g\phi T \over 4\pi} \ll {T^2\over
12}$. In other words, ${C  \phi^2\over T } \dot\phi \ll g^2\phi
\langle\chi^2\rangle_{eq}$, so the new viscosity term in Eq.
(\ref{PPPP}) is always much smaller than the usual thermal correction
to the equation of motion of the field $\phi$. This is an important
fact, which will help us to evaluate plausibility of warm inflation.

As an example, let us consider two limiting cases:

\begin{description}
\item[Case I.] $ gT\ll m$. \\
In this case $m^2\phi \gg g^2\phi\langle\chi^2\rangle_{eq} \gg {C
\phi^2\over  T } \dot\phi$, so the new viscosity term is completely
irrelevant for the description of the evolution of the field $\phi$,
as is clear from the above intuitive derivation of the equation of
motion.

\item[Case II.] $ gT\gg m$. \\
If $g\phi \gg T$, then all thermal effects disappear. If  $g\phi \ll
T$, then $m^2\phi^2 \ll g^2\phi^2\langle\chi^2\rangle_{eq} \sim g^2
\phi^2 T^2 \ll T^4$. In this case equation of state is determined by
ultrarelativistic matter, $p \approx \rho/3$, and inflation cannot
occur.

\end{description}

In conclusion,   new viscosity terms in the equation of motion of the
inflaton field appear as a correction to a correction to the usual
high temperature terms such as $g^2\phi\langle\chi^2\rangle$. These
terms appear  if one wants to write equation of motion of the scalar
field in terms of the thermal equilibrium quantities such as
$\langle\chi^2\rangle_{eq}$, which differs only very slightly from
its out-of-equilibrium value $\langle\chi^2\rangle$. If the leading
thermal effects cannot result in existence of an inflationary regime
of a new type, then the sub-subleading terms cannot do it as well.

In this investigation we did not really evaluate the viscosity term
${C \phi^2\over  T } \dot\phi$; we only   used our knowledge that
this term appeared as a correction to a correction to the term
${g^2\over 12} T^2$, so it should be always much smaller than
${g^2\over 12} T^2$. It is very instructive to study the same issue
by solving   equation (\ref{PPPP}) directly, without making any {\it
a priori} assumptions about the magnitude of the viscosity term.

Suppose one has a stage of warm inflation in the regime where $T
> g\phi$,
but   $m^2 \gg g^2T^2$ (since otherwise one either does not have
thermal corrections, or one has a noninflationary radiation dominated
universe). In this regime, and assuming that ${C \phi^2\over  T } \gg
H$,  Eq. (\ref{PPPP}) acquires the following form:
\begin{equation}\label{PPPPa}
  \ddot{\phi}+{C  \phi^2\over  T } \dot\phi +
 m^2\phi   =0.
\end{equation}
In the regime of slow rolling one can neglect the first term and
obtain a solution
\begin{equation}\label{PaPa}
\phi^2 = \phi_0^2 -   {2m^2 T t \over C},
\end{equation}
 assuming, for simplicity, that the temperature remains constant.
One can easily verify that for $gm  < T < {Cm\over g}$ and $\phi^2
> {mT\over C}$ this solution is compatible with the inequalities
$T > g\phi$, but   $m^2 \gg g^2T^2$, and the term $\ddot \phi$ could
indeed be neglected in Eq. (\ref{PPPPa}). Since it is not a usual
oscillatory regime, one could think that we have an interesting
candidate for the warm inflation regime.

However, Eq.  (\ref{PaPa}) tells us that the value of the field
$\phi$ completely changes (vanishes) within the time $\delta t =
{C\phi_0^2\over 2m^2T}$. As one can easily verify, in the regime
which we study now ($g\phi \ll T$
 and  $m^2 \gg g^2T^2$) this time  is much smaller than the
typical decay time of the $\chi$ particles:
\begin{equation}\label{decay}
\Gamma_\chi \delta t \sim \frac{g^4 C\phi_0^2 }{384\pi m^2}  \ll
\frac{  C\ }{384\pi  }\ll 1 \ .
\end{equation}
Thus the simple linear approximation  $\phi - \Delta \phi \sim \phi -
\dot\phi \Delta t$, with $\Delta t \sim \Gamma_\chi^{-1}$, which we
used for the derivation of Eq. (\ref{PPPP}) does not work in
application to our problem: The typical duration of the linear change
of the field $\phi$ is much shorter than $\Gamma_\chi^{-1}$. This
means that Eq.   (\ref{PPPP}) is incorrect in the only regime where
warm inflation could appear in this model. In our previous
qualitative analysis of warm inflation in this model we avoided this
problem because we did not used  the expression ${C \phi^2\over  T }
\dot\phi$ for the viscosity term, which we obtained using the
incorrect assumption that $\phi - \Delta \phi \sim \phi - \dot\phi
\Gamma_\chi^{-1}$. Instead of that, we used the  fact that the
viscosity term
 is always much smaller than ${g^2\over 12} T^2\phi$.

To study motion of the field $\phi$ in this model one should remember
that  the typical time when a significant change of the scalar field
occurs (such as the oscillation time $m^{-1}$) is much shorter than
$\Gamma_\chi$. Therefore the number of $\chi$ particles practically
does not change during the oscillations. To study the motion of the
field $\phi$ in this case one should evaluate $\langle\chi^2\rangle$
in the regime where $n_{\chi}$ remains constant, see e.g.
\cite{KLS97}. The behavior of the field $\phi$ in this regime is
completely different from what one could expect on the basis of Eq.
(\ref{PPPP}).

This example shows one of the many obstacles which make  it so
hard or perhaps even impossible to implement the idea of warm
inflation. In order to verify our conclusions, we performed a more
detailed investigation of this issue and studied several different
models where warm inflation could occur. Before starting its detailed
description we remind the reader that the above intuitive derivation
of equation (\ref{PPPP}) is based on the following two most
fundamental assumptions or conditions:
\begin{description}
\item[(A)] $\phi$  interacts with particles whose effective mass
is much smaller than the temperature.
\item[(B)] $\phi$ should not evolve significantly during
the relaxation time, $\Delta t$, of such particles, namely,
$|\dot{\phi}|\Delta t \ll \phi$. (Otherwise the quasi-stationary
inflationary expansion could not be sustained.)
\end{description}

A complete discussion of warm inflation is rather involved; in
addition to these two conditions many other constraints have to be
satisfied. However, as we will see in order to show that the
scenario is not
feasible in all models that we will consider it is enough to use only the
simplest constraints based on
the above two conditions. In the next section we review derivation of
the equation of motion of the scalar field in more generic cases in
terms of quantum field theory at finite temperature, where these two
conditions (A) and (B) are also indispensable.  It
 has been advocated
in the literature \cite{bgr} that this method is quite appropriate
for application to warm inflation where the scalar field presumably
moves slowly.

As we have  demonstrated above for a particular model of two scalar
fields with interaction ${g^2\over 2} \phi^2\chi^2$, however, $\phi$
does evolve significantly during the relaxation time even in its slow
roll-over regime, so that the condition (B) is not satisfied. Put
it differently, if we attempt to realize warm inflation consistently
with the derivation of the equation of motion in the form of Eq.
(\ref{PPPP}), we inevitably find that the $e$-folding number of
inflation is constrained to be much smaller than unity. In fact,
a similar result has also been obtained in \cite{bgr} for this
particular model. However, there was a hope that this result is not
generic, and several possible ways out to increase the number of
$e$-folds have  been proposed \cite{bgr}. In Sec.\ \ref{warm} we
demonstrate for several other models that the number of $e$-folds of
inflation must be much smaller than unity. As will be seen there,
this conclusion remains unchanged even when we take the proposals
of \cite{bgr} into account.

%%%%%%%%%%%%%%%%%%%%%%%%%%%%%%%%%%%%%%%%%%%%%%%%%%%%%%%%%%%%%%%%%%%%%%%%%%
\section{Effective equation of motion in the thermal bath}
\label{derivation}
%%%%%%%%%%%%%%%%%%%%%%%%%%%%%%%%%%%%%%%%%%%%%%%%%%%%%%%%%%%%%%%%%%%%%%%%%%

\indent

As a preparation to quantitative discussion which confirms the above
intuitive argument here
 we review a field theoretic approach to derive an effective
equation of motion in the thermal bath. The standard quantum field
theory, which is appropriate for evaluating the transition amplitude
from an `in' state to an `out' state for some field operator
$\vect\CO$, $\la out|\,\vect\CO\,|in \ra$, is not suitable to trace
time evolution of an expectation value in a non-equilibrium system
interacting with a thermal bath. For this purpose we should use the
in-in formalism, which was introduced
 by Schwinger
\cite{Sch} and developed by Keldysh \cite{Kel}. Here, extending
Gleiser and Ramos \cite{GR} and Yamaguchi and Yokoyama \cite{YY}, who
followed Morikawa \cite{Mor}, we derive an effective Langevin-like
equation for a coarse-grained field in the case the scalar field is
interacting with not only other scalars but also fermions.

\subsection{Non-equilibrium quantum field theory}\label{non}

\indent

Let us consider the following Lagrangian density of a singlet scalar
field $\varphi$ interacting with another scalar field $\chi$ and a
fermion $\psi$ for illustration.
\beq
   \CL = \frac12\,(\del_{\mu}\varphi)^2-\frac12\,m_{\varphi}^2\varphi^2 -
     \frac{1}{4!}\lambda\,\varphi^4
       + \frac12\,(\del_{\mu}\chi)^2-\frac12\,m_{\chi}^2\chi^2 -
     \frac{1}{2}\,g^2 \chi^2 \varphi^2
       + i\bar\psi\gamma^{\mu}\del_{\mu}\psi - m_{\psi} \bar\psi \psi
     - f\varphi\bar \psi\psi \:.
\eeq

In order to follow the time development of $\varphi$, only the
initial condition is fixed, and so the time contour in a generating
functional starting from the infinite past must run to the infinite
future without fixing the final condition and come back to the
infinite past again. The generating functional is thus given by
\bea
  Z[J,\eta,\bar\eta] &\equiv&
      Tr \lkk T\lmk \exp
        \lkk i\int_{c} (J \varphi+K \chi +\eta\psi+\bar\eta \bar\psi
          \rkk \rmk \rho\,\rkk \non \\
       &=&
          Tr \lkk T_{+}\lmk
            \exp\lkk i\int
(J_{+}\varphi_{+}+K_{+}\chi_{+}+\eta_{+}\psi_{+}
                +\bar\eta_{+}\bar\psi_{+})
                    \rkk \rmk \right. \non \\
       && \qquad \qquad \qquad \left.   \times\,
          T_{-}\lmk \exp\lkk i\int
                (J_{-}\varphi_{-}+K_{-}\chi_{-}+\eta_{-}\psi_{-}+
                    \bar\eta_{-} \bar\psi_{-})
               \rkk \rmk \rho\,\rkk \:, \non \\
\eea
where the suffix $c$ represents the closed time contour of
integration and $X_{+}$ a field component $X$ on the plus-branch
($-\infty$ to $+\infty$), $X_{-}$ that on the minus-branch ($+\infty$
to $-\infty$). The symbol $T$ represents the time ordering according
to the closed time contour, $T_{+}$ the ordinary time ordering, and
$T_{-}$ the anti-time ordering. $J, K, \eta$, and $\bar\eta$ imply
the external fields for the scalar and the Dirac fields,
respectively. In fact, each external field
$J_{+}(K_{+},\eta_{+},\bar\eta_{+})$ and
$J_{-}(K_{-},\eta_{-},\bar\eta_{-})$ is identical, but for technical
reason we treat them different and set
$J_{+}=J_{-}(K_{+}=K_{-},\eta_{+}=\eta_{-},\bar\eta_{+}=\bar\eta_{-})$
only at the end of calculation. $\rho$ is the initial density matrix.
Strictly speaking, we should couple the time development of the
expectation value of the field with that of the density matrix, which
is practically impossible. Accordingly we assume that deviation from
the equilibrium is small and use the density matrix corresponding to
the finite-temperature  state. Then the generating functional is
described by the path integral as \beq
  Z[\,J,K,\eta,\bar\eta\,]
    = \exp \biggl(\,iW[\,J,K,\eta,\bar\eta\,]\,\biggr)
     = \int_{c}\CD\varphi \int_{c}\CD\chi \int_{c}\CD\psi
        \int_{c}\CD\psi^{\ast}
        \exp{iS[\,\varphi,\chi,\psi,\bar\psi,J,K,\eta,\bar\eta\,]} \:,
\eeq where the classical action $S$ is given by \beq
  S[\,\varphi,\chi,\psi,\bar\psi,J,K,\eta,\bar\eta\,]=
              \int_{c}d^4x \lhk \CL+J(x)\varphi(x)+K(x)\chi(x)
                 +\eta(x)\psi(x)+\bar\eta(x)\bar\psi(x) \rhk \:.
\eeq As with the Euclidean-time formulation, the scalar field is
still periodic and the Dirac field anti-periodic along the imaginary
direction, now with $\varphi(t,\vect x)=\varphi(t-i\beta,\vect x)$,
$\chi(t,\vect x)=\chi(t-i\beta,\vect x)$, and $\psi(t,\vect
x)=-\psi(t-i\beta,\vect x)$.

The effective action for the scalar field is defined by the connected
generating functional as \beq
\Gamma[\phi]=W[\,J,K,\eta,\bar\eta\,]-\int_{c}d^4x J(x)\phi(x) \:,
   \label{eqn:a1effe}
\eeq where $\phi(x)=\delta W[J,K,\eta,\bar\eta] / \delta J(x)$.

We give the finite temperature  propagator before the perturbative
expansion. For the closed path, the scalar propagator has four
components.

\begin{eqnarray}
  G_{\chi}(x-x') &=& \pmatrix{
G^{F}_{\chi}(x-x') & G^{+}_{\chi}(x-x') \cr
G^{-}_{\chi}(x-x') & G^{-}_{\chi}(x-x') \cr
}\\
 \non \\
                    &\equiv&
 \pmatrix{
Tr[\,T_{+}\chi(x)\chi(x')\rho\,] & Tr[\,\chi(x')\chi(x)\rho\,]  \cr
Tr[\,\chi(x)\chi(x')\rho\,] & Tr[\,T_{-}\chi(x)\chi(x')\rho\,] \cr
}
\end{eqnarray}
Similar
formulae apply for $\varphi$ field as well.

Also, for a Dirac fermion we find
\bea S_{\psi}(x-x') &=& \pmatrix{
S^{F}_{\psi}(x-x') & S^{+}_{\psi}(x-x')  \cr
S^{-}_{\psi}(x-x') & S^{\tilde F}_{\psi}(x-x') \cr
}\\
 \non \\
                    &\equiv&
\pmatrix{
Tr[\,T_{+}\psi(x)\bar\psi(x')\rho\,] &
Tr[\,-\bar\psi(x')\psi(x)\rho\,]  \cr
Tr[\,\psi(x)\bar\psi(x')\rho\,] & Tr[\,T_{-}\psi(x)\bar\psi(x')\rho\,] \cr
}
\eea
%where
%\bea
 %  S^{F}_{\psi}(k) &=&\frac{i}{\not{k}-m_{\psi}+i\epsilon}
 %      -2\pin_{\psi}(\vect k)
 % (\not{k}+m_{\psi})\,\delta(k^2-m_{\psi}^2) \:, \non \\
 % S^{\tilde F}_{\psi}(k) &=&\frac{-i}{\not{k}-m_{\psi}-i\epsilon}
 % -2\pi n_{\psi}(\vect k)
 %(\not{k}+m_{\psi})\,\delta(k^2-m_{\psi}^2) \:, \non \\
 %S^{+}_{\psi}(k) &=& -2\pi\,
 % n_{\psi}(\vect k)\epsilon(k_{0})(\not{k}+m_{\psi})
 % \,\delta(k^2-m_{\psi}^2) \:,\non \\
 %  S^{-}_{\psi}(k) &=& 2\pi\, (1-n_{\psi}(\vect k))\epsilon(k_{0})
 % (\not{k}+m_{\psi}) % \,\delta(k^2-m_{\psi}^2)\:,
  %\eea
%with $n_{\psi}(\vect k)=(e^{\beta\omega_{\psi}(\vect k)}+1)^{-1},\,
%\omega_{\psi}(\vect k)=\sqrt{\vect k^2+m_{\psi}^2}$ \cite{pro}.

\subsection{One-loop finite temperature  effective action}\label{oneloop}

\indent The perturbative loop expansion for the effective action
$\Gamma$ can be obtained by transforming $\varphi \rightarrow
\varphi_{0}+\zeta$ where $\varphi_{0}$ is the field configuration
which extremizes the classical action $S[\,\varphi,J\,]$ and $\zeta$
is small perturbation around $\varphi_{0}$. Up to one loop order and
$\CO(\lambda^2, g^4, f^2)$, %$\Gamma$ is made up of the graphs as
depicted in Fig.\ %\ref{fig:one}. Summing up these graphs, the
effective action $\Gamma$ becomes \bea
  \Gamma[\phi_{c},\phi_{\Delta}] &= &\int d^4x \lhk
        \phi_{\Delta}(x)[\,-\Box-M^2\,]\phi_{c}(x)
        -\frac{\lambda}{4!}
           \lmk
             4\phi_{\Delta}(x)\phi_{c}^3(x)+\phi_{c}(x)\phi_{\Delta}^3(x)
           \rmk
                                           \rhk \non \\
        &&-\int d^4x\int d^4x A_{1}(x-x')
           \lkk\,
             \phi_{\Delta}(x)\phi_{c}(x)\phi_{c}^2(x')
             +\frac14 \phi_{\Delta}(x)\phi_{c}(x)\phi_{\Delta}^2(x')
           \,\rkk  \non \\
        &&-2\int d^4x\int d^4x'
              A_{2}(x-x')\phi_{\Delta}(x)\phi_{c}(x') \non \\
        &&+\frac{i}{2}\int d^4x\int d^4x'
          \biggl[\,
   B_{1}(x-x')\phi_{\Delta}(x)\phi_{\Delta}(x')\phi_{c}(x)\phi_{c}(x')
         %        \non \\
        % && \qquad \qquad \qquad \qquad
               +B_{2}(x-x')\phi_{\Delta}(x)\phi_{\Delta}(x')
          \,\,\biggr]  \:, \non \\
        &&
  \label{eqn:a3effe}
\eea where \bea
   \phi_{c} &\equiv& \frac12(\phi_{+}+\phi_{-}) \:, \\
   \phi_{\Delta} &\equiv& \phi_{+}-\phi_{-} \:, \\
   M^2 &=& m^2 + g^2 \int\frac{d^3  q}{(2\pi)^3}
             \frac{1+2n_{\chi}(\vect q)}{2\omega_{\chi}(\vect q)}
             +\frac{\lambda}{2} \int\frac{d^3  q}{(2\pi)^3}
             \frac{1+2n_{\varphi}(\vect q)}
               {2\omega_{\varphi}(\vect q)} \:, \\
   A_{1}(x-x') &=& 2g^4
                    \mbox{Im}\,[\,G^{F}_{\chi}(x-x')^2\,]
                      \,\theta(t-t')+\frac{\lambda^2}{2}
                    \mbox{Im}\,[\,G^{F}_{\varphi}(x-x')^2\,]
                      \,\theta(t-t')
         \label{eqn:A1}  \:. \\
   A_{2}(x-x') &=& f^2 \mbox{Im}\,[\,S_{\alpha\beta}^{F}(x-x')
                    S_{F}^{\beta\alpha}(x'-x)\,]\,\theta(t-t')
         \label{eqn:A2}  \:. \\
   B_{1}(x-x') &=& 2g^4
                    \mbox{Re}\,[\,G^{F}_{\chi}(x-x')^2\,]
                    +\frac{\lambda^2}{2}
                    \mbox{Re}\,[\,G^{F}_{\varphi}(x-x')^2\,]
         \label{eqn:B1}  \:. \\
   B_{2}(x-x') &=& -f^2 \mbox{Re}\,[\,S_{\alpha\beta}^{F}(x-x')
                    S_{F}^{\beta\alpha}(x'-x)\,]
         \label{eqn:B2}  \:.
         %\\
\eea

The last term of (\ref{eqn:a3effe}) gives the imaginary contribution
to the effective action $\Gamma$.   We can attribute these imaginary
terms to the functional integrals over real auxiliary fields
$\xi_{1}(x)$ and $\xi_{2}(x)$ \cite{Mor} to rewrite
(\ref{eqn:a3effe}) as \beq
  \exp (i\Gamma[\phi_{c},\phi_{\Delta}])=\int\CD\xi_{1} \int\CD\xi_{2}
              P_{1}[\xi_{1}]P_{2}[\xi_{2}]\exp\lhk
                 iS_{eff}[\,\phi_{c},\phi_{\Delta},\xi_{1},\xi_{2}\,]\rhk
                    \:,
\eeq where \beq
  S_{\eff}[\,\phi_{c},\phi_{\Delta},\xi_{1},\xi_{2}\,] \equiv
                 \mbox{Re}\Gamma
                  +\int d^4x[\,\xi_{1}(x)\phi_{c}(x)\phi_{\Delta}(x)
                    +\xi_{2}(x)\phi_{\Delta}(x)\,] \:.
 \label{eqn:a3cla}
\eeq Here $\xi_1(x)$ and $\xi_2(x)$ are random Gaussian fields with
the probability distribution functional \beq
  P_{i}[\xi_{i}] = \CN_{i} \exp \lkk\, - \frac12\int d^4x \int d^4x'
                      \xi_{i}(x) B_{i}^{-1}(x-x')\xi_{i}(x')\,\rkk
                       \qquad (i = 1, 2)    \:.
\eeq where $\CN_{i}$ is a normalization factor. They induce random
noise terms in the effective equation of motion of  $\phi$ as a
result of the interactions with the thermal bath.

\subsection{Equation of motion}\label{motion}
\label{sub:lan}

\indent Applying the variational principle to $S_{\eff}$, we obtain
the equation of motion for $\phi_{c}$. \beq
  \frac{\delta S_{\eff}[\,\phi_{c},\phi_{\Delta},\xi_{1},\xi_{2}\,]}
        {\delta \phi_{\Delta}}
   \biggl. \biggr|_{\phi_{\Delta}=0}
      =0 \:.
\eeq
 From (\ref{eqn:a3cla}) and (\ref{eqn:a3effe}), it reads
\bea
  (\,\Box\! &+&\! M^2\,)\,\phi_{c}(x)
          +\frac{\lambda}{3!}\phi_{c}^3(x) \non \\
      &+& \phi_{c}(x)\int d^3  {x'}\int_{-\infty}^{t}dt'
               A_{1}(x-x')\phi_{c}^2(x')
           +2\int d^3
            {x'}\int_{-\infty}^{t}dt'A_{2}(x-x')\phi_{c}(x')
             \non \\
           && \qquad \qquad \qquad \qquad \qquad \qquad \qquad
           =\phi_{c}(x)\xi_{1}(x)+\xi_{2}(x) \:,
  \label{eqn:a4eqm}
\eea and \beq
  \la\,\xi_{i}(x)\xi_{i}(x')\,\ra = B_{i}(x-x') \:.
  \label{eqn:corre}
\eeq Though $A_{1}$ and $B_{1}$ has two contributions from $\chi$ and
$\varphi$ fields, they have the same properties except for the values
of coefficients and masses. For the moment, we consider only the
contribution from $\varphi$ field for simplicity and omit the suffix
$c$. The right hand side of (\ref{eqn:a4eqm}) are the noise terms,
while the last two terms of the left hand side are combination of
dissipation term and one-loop correction to the classical equation of
motion which would reduce to a part of the derivative of the
effective potential, $V'_{\eff}(\phi)$, if we restricted $\phi(x')$
to be a constant in space and time.

The above equation (\ref{eqn:a4eqm}) is an extension of equation
(3.2) of Gleiser and Ramos \cite{GR} in that we have incorporated not
only self-interaction but also interactions with a boson $\chi$ and a
fermion $\psi$. It is nonlocal in space and time. The spatial
nonlocality does not bring any difficulty here since the scalar field
is presumably nearly homogeneous during inflation.  So we only need
to consider contributions with zero external momentum, that is, we
can put $\phi_c(\vex',t')=\phi_c(\vex,t')$ in the integrand.

 With this approximation the correlation function
of the bosonic noise (\ref{eqn:B1}) with $g=0$, for example, becomes
\bea
  \la\,\xi_{1}(x)\xi_{1}(x')\,\ra
         &\Rightarrow& \left.
           \frac{\lambda^2}{2}
            \int\frac{d^3  k}{(2\pi)^3}
             e^{i\vect k\cdot(\vect x-\vect x')}
              \int\frac{d^3  q}{(2\pi)^3}
                \mbox{Re}\,[\,G^{F}_{\phi}(\vect q,t-t')
                   G^{F}_{\phi}(\vect{q-k},t-t')\,]
                    \right|_{\vect{k}=\vect{0}}
                       \non  \\
         &=& \frac{\lambda^2}{2} \delta^3(\vect x-\vect x')
               \int\frac{d^3  q}{(2\pi)^3}
                \mbox{Re}\,[\,G^{F}_{\phi}(\vect q,t-t')\,]^2
                 \:.
             \label{eqn:homas}
\eea We thus obtain spatially uncorrelated noise.  For the scalar
field averaged over a volume $V$, the amplitude of noise reduces in
proportion to $V^{-1/2}$, which implies that noise term should be
omitted in the equation of motion of spatially-averaged homogeneous
field.  On the other hand, the temporal nonlocality is very important
in deriving the dissipation term as seen below.

\subsection{Dissipation term}\label{diss}

\indent

The equation of motion (\ref{eqn:a4eqm}) derived above has
contributions representing the dissipative effect in the last two
terms of the left hand side. Since these terms are nonlocal in time,
in order to extract local terms proportional to $\dot\phi$ one should
assume that the field changes adiabatically \cite{Mor} \cite{GR} , or
put

\beq
  \phi^n(\vect x',t') \simeq \phi^n(\vect x',t)
   + n(t'-t) \phi^{n-1}(\vect x',t) \dot\phi(\vect x',t) \
  \label{eqn:adia}
\eeq
\noindent in the integrand of (\ref{eqn:a4eqm}). Then these terms
will read \bea
 &&\phi\int d^3x'\int_{-\infty}^t dt' A_1(x-x')\phi^2(x')+
 2\int d^3x'\int_{-\infty}^t dt' A_2(x-x')\phi(x') \non \\
 &=& \phi^3(t)\int d^3x'\int_{-\infty}^t dt' A_1(x-x')
 + 2\phi(t)\int d^3x'\int_{-\infty}^t dt'A_2(x-x') \non \\
 &&+2\phi^2(t)\dot{\phi}(t)\int d^3x'\int_{-\infty}^t
 dt'(t'-t)A_1(x-x')
 +2\dot{\phi}(t)\int d^3x'\int_{-\infty}^t dt'(t'-t)A_2(x-x').
\eea The last two terms are dissipation terms. They would
vanish if we used bare propagators, as a manifestation of the fact
that the dissipative effect is intrinsically a non-perturbative
phenomenon and cannot be investigated from the perturbation theory
\cite{BVHLS}. In order to obtain a finite result we should use full
 ``dressed" propagators instead \cite{Mor,GR,bgr}.

The viscosity from scalar interactions have been fully investigated
in \cite{GR} and \cite{bgr}, so we simply quote their results here.
The full propagator of $\phi$ reads \beqa
  G_\phi^F(\vek,t-t')&=&\int\frac{d^3k}{(2\pi)^3}e^{i\vek(\vex-\vex')}
  \non \\
  &=&\frac{i}{2\omega_\phi}\lhk\lkk
  1+n_B(\omega_\phi-i\Gamma_\phi)\rkk
  e^{-i(\omega_\phi-i\Gamma_\phi)|t-t'|}
  +n_B(\omega_\phi+i\Gamma_\phi)e^{i(\omega_\phi+i\Gamma_\phi)|t-t'|}
  \rnk,
\eeqa where $n_B(\omega_\phi)=\lmk e^{\beta\omega_\phi}-1\rmk^{-1}$
and $\omega_\phi=\sqrt{\vek^2+m^2_{\phi T}}$ with $m_{\phi T}$ being
the finite-temperature effective mass of $\phi$.  Here $\Gamma_\phi$
is the decay width related with the imaginary part of the self
energy, $\Sigma$, as \beq
\Gamma_\phi=-\frac{\mbox{Im}\Sigma_\phi}{2\omega_\phi}\simeq
\frac{\lambda^2T^2}{1536\pi\omega_\phi}, \eeq in the limit $T \gg
m_{\phi T}$. We find the contribution of the self-interaction to the
viscosity \beqa
  \lambda^2\phi^2(t)\dot{\phi}(t)\int d^3x'\int_{-\infty}^t
 dt'(t'-t)\mbox{Im}\,[\,G^{F}_{\phi}(x-x')^2\,]
 &\simeq&\frac{\lambda^2}{8}\phi^2\dot{\phi}\beta
 \int\frac{d^3k}{(2\pi)^3}\frac{n_B\lmk
 1+n_B\rmk}{\omega_\phi^2\Gamma_\phi} \label{bosevisc} \\
 &\simeq& \frac{96}{\pi
 T}\ln\lmk\frac{T}{m_{\phi T}}\rmk \phi^2\dot{\phi}, \non
\eeqa in the high temperature limit and in the case $g=f=0$. Note
that the above expression was first obtained by Hosoya and Sakagami
\cite{HS} by a different method which is intuitively more appealing
as discussed in Sec.\ \ref{intuit}.

The contribution from the interaction with $\chi$ can be obtained
similarly.  Assuming $\chi$ has no self-interaction its width is
given by \beq
  \Gamma_\chi \simeq \frac{g^4 T}{192\pi}
\eeq in the high-temperature limit and the relevant part of the
viscosity terms is given by \beq
  4g^4\phi^2(t)\dot{\phi}(t)\int d^3x'\int_{-\infty}^t
 dt'(t'-t)\mbox{Im}\,[\,G^{F}_{\chi}(x-x')^2\,]
 \simeq \frac{48}{\pi T}\ln\lmk\frac{T}{m_{\chi
 T}}\rmk\phi^2\dot{\phi}.
\eeq In both cases the viscosity terms due to scalar interactions are
of the form $\phi^2\dot{\phi}/T$ \cite{bgr}.

On the other hand, the dissipation due to Yukawa interaction is
calculated in the Appendix.  In the high temperature limit $T \gg
m_{\psi T},~m_{\phi T}$ we find \beqa
  2f^2\dot{\phi}\int d^3x'\int_{-\infty}^t dt' (t'-t)\mbox{Im}
  \,[\,S_{\alpha\beta}^{F}(x-x')
                    S_{F}^{\beta\alpha}(x'-x)\,]\,
  &\simeq& \frac{288}{\pi^3}\zeta (3) T\dot{\phi} \label{fermivisc}
 \simeq
  11T\dot{\phi}. \non
\eeqa

In both cases the viscosity coefficients would be exponentially
suppressed and would not play any important role if the
high-temperature conditions were not satisfied.

\section{Feasibility of warm inflation}  \label{warm}

In this section we study if the warm inflation driven by the
viscosity term is possible in the case the viscosity term in the
effective equation of motion is most effective, namely, in the
high-temperature limit when the viscosity coefficient is given either
by (\ref{bosevisc}) or (\ref{fermivisc}) depending on the
interaction.  We consider these cases separately for both new
\cite{newinf} and chaotic \cite{chaoinf} inflation scenarios.

Before analyzing specific models we formulate generic conditions to
satisfy.  We are interested in the new possibility that the
slow-rollover inflation is realized due to the thermal viscosity
(\ref{bosevisc}) or (\ref{fermivisc}), so the effective equation of
motion should read \beq
  C_v\dot{\phi}=-V'[\phi],  \label{slowroll}
\eeq that is, we require $  C_v \gg 3H$. The inflaton's energy
released through this viscosity term presumably goes to radiation,
whose energy density, $\rho_r$, satisfies \beq
  \frac{d\rho_r}{dt}=-4H\rho_r +C_v\dot{\phi}^2.  \label{rad}
\eeq Since $C_v$ strongly depends on the radiation temperature,
$\rho_r$ should not change too rapidly in time in order to sustain
quasi-stationary inflation.  So the creation term in (\ref{rad})
should balance the redshift term.  As a result we find \beq
  \rho_r=\frac{\pi^2g_*}{30}T^4 \simeq
  \frac{C_v\dot{\phi}^2}{4H}, \label{radiation}
\eeq which gives the radiation temperature as a function of $\phi$.
In contrast to \cite{bgr}, where the temperature has been fixed to
its initial value, we perform a consistent analysis by using the
value of $T$ calculated from (\ref{radiation}), which would help to
increase the viscosity from bosonic interactions \cite{bgr}.
 Here $g_*$ is the total
effective number of relativistic degrees of freedom.  We normalize it
by 150 and denote $ g_*/150 \equiv \gn$ below. In order that the
universe is inflating the potential energy density should dominate
over $\rho_r$ and the kinetic energy density.

Finally a number of conditions must be satisfied to justify the
derivation of the effective equation of motion. In order that the
radiation created from the inflaton thermalizes sufficiently rapidly,
we need \beq \Gamma_\phi\simeq \frac{\lambda^2T}{1536\pi} \gg H,~~~
 \Gamma_\chi\simeq \frac{g^4 T}{192\pi} \gg H,~~~
 \Gamma_\psi\simeq \frac{\pi}{64}\frac{f^2m_{\psi T}^2}{T} \gg H,
 \label{gammah}
\eeq depending on the nature of interaction. In addition, the
adiabatic conditions \beq \Gamma_\phi,~\Gamma_\chi,~
 \Gamma_\psi \gg \left|\frac{\dot{\phi}}{\phi}\right|, \label{gammaphi}
\eeq must be fulfilled so that the viscosity term is proportional to
$\dot{\phi}$. The inequalities (\ref{gammah}) and (\ref{gammaphi})
are essential to realize quasi-stationary state of inflationary
expansion. Finally, but most importantly, the high-temperature
conditions $T \gg m_{\phi T},~m_{\chi T},~m_{\psi T}$ must also be
satisfied, otherwise the viscosity term would be exponentially small
and our entire discussion would break down.

Since the failure of the {\bf Case II} in Sec.\ \ref{intuit} is
evident, we here consider the {\bf Case I}, where finite-temperature
correction to the effective potential is sub-dominant but still the
viscosity term appears sizable at first glance (until we convince
ourselves it is not, through the intuitive argument in Sec.\
\ref{intuit}.) Thus we can make the following list of the
inequalities to be satisfied to realize the desired scenario.
\begin{description}
  \item[(i)] $C_v \gg 3H$.
  \item[(ii)] $V[\phi] \gg \frac{1}{2}\dot{\phi}^2$.
  \item[(iii)] $V[\phi] \gg \rho_r$.
  \item[(iv)] $\Gamma_\phi \gg H$, $\Gamma_\chi \gg H$, or
    $\Gamma_\psi \gg H$.
  \item[(v)] $\Gamma_\phi \gg \displaystyle{\left|\frac{\dot{\phi}}
  {\phi}\right|}$, $\Gamma_\chi \gg \displaystyle{\left|\frac{\dot{\phi}}
  {\phi}\right|}$,
  or $\Gamma_\psi \gg
\displaystyle{\left|\frac{\dot{\phi}}{\phi}\right|}$.
  \item[(vi)] $T \gg m_{\phi T}$.
  \item[(vii)] $T \gg m_{\chi T}$ or $T \gg m_{\psi T}$.
  \item[(viii)] Finite-temperature correction to the effective
  potential being sub-dominant.
\end{description}
Inequalities (v) are nothing but the condition (B) in
Sec.\ \ref{intuit}, and (vi) and (vii) stand for the condition (A).
As will be seen below, we can essentially rule out all the models we
consider only in terms of these two conditions.

In addition to the constraints listed above, there is  a  constraint  
following
from the investigation of density
perturbations produced during warm inflation.

The standard expression for density perturbations produced during
inflation in the cold-matter dominated universe is \cite{book}
\begin{equation}\label{p1}
{\delta\rho\over \rho} = {6\over 5}\, {H\delta\phi \over \dot\phi}
\end{equation}
If one uses the standard estimate for inflationary perturbations
$\delta\phi \sim {H\over 2\pi}$, one gets
\begin{equation}\label{p2}
{\delta\rho\over \rho} = {6\over 5}\, {H^2 \over  2 \pi\dot\phi}
\end{equation}

The first of these two equations   holds for the warm inflation as
well. However, the second one should be   modified because the
amplitude of fluctuations during warm inflation is greater than
${H\over 2\pi}$ for two different  reasons.

First of all, the wavelength of perturbations of a field with mass
$m^2 \ll H^2$ which freeze during inflation usually is $O(H^{-1})$
because of the friction term $3H\dot\phi$. This allows one to make
the standard estimate $\delta\phi \sim {H\over 2\pi}$ by calculating
the amplitude of vacuum fluctuations with the wavelength $O(H^{-1})$.
However, during warm inflation the amplitude of perturbations of
scalar fields with momenta $k \ll T$ is enhanced by the factor $\sim
\sqrt {T\over k}$ because of the additional contribution of thermal
fluctuations. This leads to an estimate $\delta\phi \sim  \sqrt
{3HT\over 4\pi}$ \cite{perturb}.

But this is not the only effect which
should be taken into account. Indeed, during warm inflation the
friction term is $(C_v+3H)\dot \phi$, and it is assumed that $C_v \gg
3H$. As a result, it may happen that the fluctuations of
  the scalar field  freeze  not when their wavelength approaches
$H^{-1}$, but much earlier,  when their amplitude can be much
greater. This may
lead to an additional increase of the magnitude of   perturbations
of scalar
fields produced during inflation.

We will not
perform here a  detailed evaluation of density perturbations in
warm inflation because we did not find any model where this scenario
can be realized.
Indeed,  we will be able to rule out warm inflation in all models which we
will consider even without  using the theory of density
perturbations. However,
one
should keep in mind the necessity to study constraints based on the
theory of
density perturbations,
because usually these constraints lead  to
the strongest restrictions on the structure of inflationary models.

\subsection{Chaotic inflation with viscosity from bosonic
interaction}

Here we consider chaotic-type inflation with a potential
$V[\phi]=\frac{\lambda}{4!}\phi^4$ and the viscosity arising from
self-interaction or other bosonic interactions proportional to
$\phi^2$, such as $\frac{1}{2}g^2\phi^2\chi^2$. In the usual chaotic
inflation slow-roll of the inflaton is realized due to the Hubble
friction and it is effective only when $\phi \gsim 0.3\mpl$, but if
thermal viscosity discussed above is effective, we might have
inflation with much smaller $\phi$.

In this case the viscosity coefficient and the Hubble parameter are
 given, respectively, by
\beq
   C_v=\frac{96c_\phi}{\pi
   T}\phi^2,~~~~~H=\frac{\sqrt{\pi\lambda}}
   {3\mpl}\phi^2.
\eeq Here $c_\phi=\ln(T/m_{\phi T})$ if $\phi$ has self-interaction
only.  But $c_\phi$ can be increased if it interacts with additional
scalar fields $\chi_i$'s through the interaction \beq
  \CL_{\rm int}=-\sum_j\frac{1}{2}g_j^2\phi^2\chi_j^2.
\eeq Then $C_v$ is given by \beq
  C_v=\frac{96}{\pi T}\lkk \frac{\lambda^2}{\lambda^2+\sum_j
  g_j^4}\ln\lmk\frac{T}{m_{\phi T}}\rmk
  +\sum_j\frac{1}{2}\ln\lmk\frac{T}{m_{\chi_j T}}\rmk\rkk,
\eeq where $m_{\chi_j T} \ll T$ is the finite-temperature effective
mass of $\chi_j$ field \cite{bgr}.  For simplicity, we identify
$c_\phi (\gg 1)$ with the number of $\chi_j$ fields, which
corresponds to the case $\ln (T/m_{\chi_j T})=2$.  We also assume all
the coupling constants $g_j$ take the same value $g_j\equiv g$.

 From (\ref{slowroll}) and (\ref{radiation}) the temperature is
given by \beq
  T=1.6\times 10^{-2}\gn^{-\frac{1}{3}}c_\phi^{-\frac{1}{3}}
  \lambda^{\frac{1}{2}}\mpl^{\frac{1}{3}}
  \phi^{\frac{2}{3}}.
\eeq We can then convert the inequalities (i)--(vi) to the
constraints on the range of $\phi$ and on other model parameters as
follows. \indent
\begin{description}
\item[(i)]  $\phi \ll 3.4\times 10^4\gn^{2}
  c_\phi^2\lambda^{-\frac{3}{2}}\mpl$.
\item[(ii)] $\phi \gg 2.9\times 10^{-11}\gn^{-1}
  c_\phi^{-4}\lambda^3\mpl$.
\item[(iii)] $\phi \gg 1.3\times 10^{-3}\gn^{-\frac{1}{4}}
  c_\phi^{-1}\lambda^{\frac{3}{4}}\mpl$.
\item[(iva)] $\phi \ll 1.2\times 10^{-4}\gn^{-\frac{1}{4}}
  c_\phi^{-\frac{1}{4}}\lambda^{\frac{3}{2}}\mpl$.
\item[(ivb)] $\phi \ll 5.5\times 10^{-4}\gn^{-\frac{1}{4}}
  c_\phi^{-\frac{1}{4}}g^3$.
\item[(va)] $\lambda \gg 26 c_\phi^{-1}$.
\item[(vb)] $\lambda g^{-4} \ll 0.30c_\phi$
\item[(vi)] $\phi \ll 1.2\times 10^{-5}\gn^{-1}c_\phi^{-1}\mpl$.
\item[(vii)] $\phi \ll 4.3\times 10^{-6}\gn^{-1}c_\phi^{-1}
  \lambda^{\frac{3}{2}}g^{-3}$.
\item[(viii)] $\lambda \gsim 3.8\times 10^{-2}c_\phi g^4$.
\end{description}
The last condition is from the requirement that radiative correction
due to $\chi$ does not change $\lambda$.

The $e$-folding number of warm inflation, if any, is calculated as
\beq
  N\equiv
  \int_{\phi_i}^{\phi_f}H\frac{d\phi}{\dot{\phi}}
  =5.1\times 10^3\gn^{\frac{1}{3}}c_\phi^{\frac{4}{3}}\lambda^{-1}
   \mpl^{-\frac{4}{3}}
   \lmk\phi_i^{\frac{4}{3}}-\phi_f^{\frac{4}{3}}\rmk, \label{cbn}
\eeq where $\phi_i$ and $\phi_f$ are upper and lower bounds of $\phi$
that satisfy all the above inequalities.

Using (vii) and (vb) in (\ref{cbn}) we find \beq
  N \ll 3.6\times 10^{-4}\gn^{-1}\lambda g^{-4} \ll 1.1\times
  10^{-4}\gn^{-1}c_\phi.
\eeq Since $\chi$'s also contribute to $g_*$, $\gn^{-1}c_\phi$ cannot
be larger than 150 and it is apparent that $N$ cannot even exceed
unity. Hence we cannot realize warm inflation in this model. This
conclusion is independent of the simplification $\ln(T/m_{\chi
T})=2$, for the above constraint on $N$ is primarily due to the
condition (vii) or $T \gg m_{\chi T}$, and the fact that this
condition is hardly satisfied implies we have overestimated the
duration of warm inflation.

Next for completeness we consider the potential $
V[\phi]=\frac{1}{2}m^2\phi^2+\frac{\lambda}{4!}\phi^4$ with $m^2 \gg
\lambda\phi^2/12 $.  In this case the temperature is given by \beq
   T=4.3\times 10^{-2}\gn^{-\frac{1}{3}}
   c_\phi^{-\frac{1}{3}}m\mpl^{\frac{1}{3}}\phi^{-\frac{1}{3}},
\eeq and the inequalities read
\begin{description}
\item[(i)]  $\phi \gg 2.8\times 10^{-2}\gn^{-\frac{1}{4}}
   c_\phi^{-1}m^{\frac{3}{4}}\mpl^{\frac{1}{4}}$.
\item[(ii)] $\phi \gg 6.0\times 10^{-2}\gn^{-\frac{1}{7}}
   c_\phi^{-\frac{4}{7}}m^{\frac{6}{7}}\mpl^{\frac{1}{7}}$.
\item[(iii)] $\phi \gg 9.1\times 10^{-2}\gn^{-\frac{1}{10}}
   c_\phi^{-\frac{2}{5}}m^{\frac{3}{5}}\mpl^{\frac{2}{5}}$.
\item[(iva)] $\phi \ll 9.5\times 10^{-5}\gn^{-\frac{1}{4}}
   c_\phi^{-\frac{1}{4}}\lambda^{\frac{3}{2}}\mpl$.
\item[(ivb)] $\phi \ll 4.5\times 10^{-4}\gn^{-\frac{1}{4}}
   c_\phi^{-\frac{1}{4}}g^3\mpl$.
\item[(va)] $\phi \gg 13c_\phi^{-\frac{1}{2}}\lambda^{-1}m$.
\item[(vb)] $\phi \gg 4.4c_\phi^{-\frac{1}{2}}g^{-2}m$.
\item[(vi)] $\phi \ll 8.0\times 10^{-5}\gn^{-1}c_\phi^{-1}\mpl$.
\item[(vii)] $\phi \ll 9.4\times 10^{-2}\gn^{-\frac{1}{4}}
c_\phi^{-\frac{1}{4}}g^{-\frac{3}{4}}m^{\frac{3}{4}}\mpl^{\frac{1}{4}}$.
\item[(viii)] $\phi < \sqrt{12}\lambda^{-\frac{1}{2}}m$.
\end{description}
The number of $e$-folds of inflation is formally given by \beq
  N=\int_{\phi_i}^{\phi_f}H\frac{d\phi}{\dot{\phi}}
  =4.4\times 10^2m^{-2}\mpl^{-\frac{4}{3}}\lmk
  \phi_i^{\frac{10}{3}} - \phi_f^{\frac{10}{3}}\rmk.  \label{cmn}
\eeq
 From (vb) and (vii) we find $m \ll 2.1\times 10^{-7}\gn^{-1}c_\phi
g^5\mpl$.  Then inequality (vii) reads $\phi \ll 9.2\times
10^{-7}\gn^{-1}c_\phi^{\frac{1}{2}}g^3\mpl$. Using these inequalities
in (\ref{cmn}) we find \beq
  N \ll 7.3\times 10^{-5}\gn^{-\frac{4}{3}}c_\phi^{-\frac{1}{3}}.
\eeq Thus no matter how many scalar fields are interacting with the
inflaton we cannot find warm inflation solution.  Note that the above
constraint on the $e$-folding number primarily comes from the
condition (vii) or $T \gg m_{\chi T}$.  This means that if we had
used the correct value of $\ln(T/m_{\chi T})$ instead of putting it
to be 2, the viscosity term would have been smaller and warm
inflation would have been even more unlikely. So we can justify our
simplification.

 Therefore the conclusion that the number of $e$-folds of warm inflation
with  chaotic-type potentials is constrained to be much smaller than
unity \cite{bgr} remains unchanged even when we use the consistent
value of the cosmic temperature obtained from (\ref{radiation}),
rather than fixing it to its initial value \cite{bgr}.

 Note that these considerations look different from the arguments
used in the Sec. \ref{intuit}, but they lead to the same final
conclusion. Now let us see how one can reach the same conclusions
directly, using the arguments of Sec. \ref{intuit}. We will assume
for simplicity that the field $\phi$ interacts with one field $\chi$,
and $\lambda \ll g^2$. In this theory Eq. (\ref{PPPP})
 looks as follows:

\begin{equation}\label{PPPPPP}
 \ddot{\phi}+C_v \dot\phi +
3H\dot{\phi}+m^2\phi
+\lambda\phi^3+g^2\phi\langle\chi^2\rangle_{eq}=0.
\end{equation}

Let us consider two limiting cases:

\begin{description}
\item[Case I.] $ g^2T^2\ll \max\, \{m^2, \lambda\phi^2 \}$. \\
In this case $m^2\phi + \lambda \phi^3  \gg
g^2\phi\langle\chi^2\rangle_{eq} \gg C_v \dot\phi$, so the new
viscosity term is completely irrelevant for the description of the
evolution of the field $\phi$, as is clear from the   intuitive
derivation of the equation of motion in Sec. \ref{intuit}.

\item[Case II.] $ g^2T^2\gg \max\, \{m^2, \lambda\phi^2 \}$. \\
If $g\phi \gg T$, then all thermal effects disappear. If  $g\phi \ll
T$, then $m^2\phi^2/2 +   \lambda \phi^4/4 \ll
g^2\phi^2\langle\chi^2\rangle_{eq} \sim g^2 \phi^2 T^2 \ll T^4$. In
this case equation of state is determined by ultrarelativistic
matter, $p \approx \rho/3$, and inflation cannot occur.
\end{description}

\subsection{Chaotic inflation with viscosity from Yukawa
interaction}

We next study the case the dominant contribution of viscosity arises
from Yukawa interaction, so that we find $C_v \simeq 11c_\psi T$
where $c_\psi=1$ in the case $\phi$ has a Yukawa coupling to one
species of Dirac fermion, but it can be larger if $\phi$ interacts
with more fermions.  We identify $c_\psi$ with the number of fermion
species interacting with $\phi$ with the universal coupling strength
$f$.

First we consider the case inflation is driven by a quartic potential
$V[\phi]=\frac{\lambda}{4!}\phi^4$.  The temperature is given by \beq
  T=0.12\gn^{-\frac{1}{5}} c_\psi^{-\frac{1}{5}}\lambda^{\frac{3}{10}}
  \phi^{\frac{4}{5}}\mpl^{\frac{1}{5}}.
\eeq The following inequalities must be simultaneously satisfied.
\begin{description}
\item[(i)]  $\phi \ll 0.78\gn^{-\frac{1}{6}}
  c_\psi^{\frac{2}{3}}\lambda^{-\frac{1}{6}}\mpl$.
\item[(ii)] $\phi \ll 62\gn^{-\frac{1}{5}}c_\psi^{4}\lambda^{-1}\mpl$.
\item[(iii)] $\phi \gg 0.15\gn^{\frac{1}{4}}
  c_\psi^{-1}\lambda^{\frac{1}{4}}\mpl$.
\item[(iv)] $\phi \ll 0.63\gn^{\frac{1}{4}}
  c_\psi^{\frac{1}{4}}f^5\lambda^{-1}\mpl$.
\item[(v)] $f^4 \gg 0.31c_\psi^{-1}\lambda$.
\item[(vi)] $\phi \ll 1.4\times
10^{-4}\gn^{-1}c_\psi^{-1}\lambda^{-1}\mpl$.
\item[(vii)]$\phi \ll 2.5\times 10^{-5}\gn^{-1}c_\psi^{-1}
     f^{-5}\lambda^{\frac{3}{2}}\mpl$.
\item[(viii)] $\phi \gg 1.6\times 10^{-6}\gn^{-1}
     c_\psi^{\frac{3}{2}}f^5\lambda^{-1}$.
\end{description}
The last inequality comes from the condition that the
finite-temperature correction to the effective potential is small,
that is, $c_\psi f^2T^2/6 \ll \lambda\phi^2/2$.

The number of $e$-folds of inflation is expressed by \beq
  N\equiv
  \int_{\phi_i}^{\phi_f}H\frac{d\phi}{\dot{\phi}}
  =5.8\gn^{-\frac{1}{5}}c_\psi^{\frac{4}{5}}\lambda^{-\frac{1}{5}}
  \mpl^{-\frac{4}{5}}
  \lmk \phi_i^{\frac{4}{5}}-\phi_f^{\frac{4}{5}}\rmk.  \label{clfn}
\eeq Using (v) in (vii) we find $\phi \ll 1.1\times 10^{-4}
\gn^{-1}c_\psi^{\frac{1}{4}}\lambda^{\frac{1}{4}}\mpl$. Inserting
this limit to $\phi_i$ in (\ref{clfn}) we obtain an upper bound on
$N$ as $N \ll 3.9\times 10^{-3}\gn^{-1}c_\psi$.  Since
$\gn^{-1}c_\psi$ cannot be larger than $300/7$ we can conclude warm
inflation is impossible here. In fact, we can explicitly show that
there is no open parameter region that satisfy all the inequalities
even in the case $\gn^{-1}c_\psi$ is maximal.

Next we analyze inflation driven by the mass term
$V[\phi]=\frac{1}{2}m^2\phi^2$, in which the temperature is given by
\beq
  T=0.19\gn^{-\frac{1}{5}}
  c_\psi^{-\frac{1}{5}}m^{\frac{3}{5}}\mpl^{\frac{1}{5}}
  \phi^{\frac{1}{5}}.
\eeq The following inequalities must be satisfied for successful warm
inflation.
\begin{description}
\item[(i)]  $\phi \ll 0.26\gn^{-\frac{1}{4}}
   c_\psi m^{-\frac{1}{2}}\mpl^{\frac{3}{2}}$.
\item[(ii)] $\phi \gg 0.025\gn c_\psi^{-4}m^{2}\mpl^{-1}$.
\item[(iii)] $\phi \gg 0.17\gn^{\frac{1}{6}}
   c_\psi^{-\frac{2}{3}}m^{\frac{1}{3}}\mpl^{\frac{2}{3}}$.
\item[(iv)] $\phi \gg 13\gn^{-\frac{1}{4}}
   c_\psi^{-\frac{1}{4}}f^{-5}m^2\mpl^{-1}$.
\item[(v)] $\phi \gg 1.4c_\psi^{-\frac{1}{2}}f^{-2}m$.
\item[(vi)] $\phi \gg 4.2\times 10^{3}\gn c_\psi m^2\mpl^{-1}$.
\item[(vii)] $\phi \ll 0.13\gn^{-\frac{1}{4}}
   c_\psi^{-\frac{1}{4}}f^{-\frac{5}{4}}
   m^{\frac{3}{4}}\mpl^{\frac{1}{4}}$.
\item[(viii)] $\phi < 3.6\times 10^5 \gn
c_\psi^{-\frac{3}{2}}f^{-5}m^2\mpl^{-1}.$
\end{description}
The last inequality is from the requirement $m^2 \gg
\frac{c_\psi}{6}f^2T^2$.

It is not impossible to find the values of model parameters that
satisfy all the above inequalities.  This does not imply, however,
that warm inflation is feasible in this model because the $e$-folding
number of inflation, \beq
  N\equiv
  \int_{\phi_i}^{\phi_f}H\frac{d\phi}{\dot{\phi}}
  =3.6\gn^{-\frac{1}{5}}c_\psi^{\frac{4}{5}}m^{-\frac{2}{5}}
  \mpl^{-\frac{4}{5}}
  \lmk \phi_i^{\frac{6}{5}} - \phi_f^{\frac{6}{5}}\rmk,  \label{nmf}
\eeq turns out to be smaller than unity as seen below, where $\phi_i$
and $\phi_f$ are upper and lower bounds satisfying all the
inequalities (i) through (viii) as before.

Inserting the upper bound (vii) to $\phi_i$ in (\ref{nmf}), we find
\beq
  N \ll 0.31\gn^{\frac{1}{2}}c_\psi^{\frac{1}{2}}f^{-\frac{3}{2}}
  m^{\frac{1}{2}}\mpl^{-\frac{1}{2}}.  \label{nmf2}
\eeq
 From the consistency between (v) and (vii)
we find $m \ll 7.4\times 10^{-5}\gn^{-1}c_\psi f^3\mpl$, which,
together with (\ref{nmf2}), implies $N \ll 2.7\times
10^{-3}\gn^{-1}c_\psi$.  Since $\gn^{-1}c_\psi$ cannot exceed $300/7$
we can conclude warm inflation is impossible in this model, too.

\subsection{New inflation}

Next we consider new inflation driven by the potential \beq
  V[\phi]=\frac{\lambda}{4}\lmk \phi^2-\frac{m^2}{\lambda}\rmk^2.
\eeq Inflation is possible only for \beq
  \phi \ll \lambda^{-\frac{1}{2}}m.  \label{phiup}
\eeq

First we analyze the case viscosity is dominated by bosonic
interaction.  The temperature is constant in this case: \beq
  T=4.8\times 10^{-2}\gn^{-\frac{1}{3}}
  c_\phi^{-\frac{1}{3}}\lambda^{\frac{1}{6}}m^{\frac{2}{3}}
  \mpl^{\frac{1}{3}}.
\eeq We find the following inequalities.
\begin{description}
\item[(i)]  $ \phi \gg 8.3\times 10^{-2}\gn^{-\frac{1}{6}}
   c_\phi^{-\frac{2}{3}}\lambda^{-\frac{1}{6}}
   m^{\frac{4}{3}}\mpl^{-\frac{1}{3}}.$
\item[(ii)] $\phi \gg 2.2\times 10^{-3}\gn^{-\frac{1}{3}}
   c_\phi^{\frac{2}{3}}\lambda^{\frac{2}{3}}m^{\frac{2}{3}}
   \mpl^{\frac{1}{3}}$.
\item[(iii)] $m \gg 6.0\times 10^{-3}\gn^{-\frac{1}{4}}
   c_\phi^{-1}\lambda^{\frac{1}{2}}\mpl$.
\item[(iva)] $m \ll 1.3\times 10^{-4}\gn^{-\frac{1}{4}}
   c_\phi^{-\frac{1}{4}}\lambda^{2}\mpl$.
\item[(ivb)] $m \ll 6.4\times 10^{-4}c_\phi^{-\frac{1}{4}}
  \gn^{-\frac{1}{4}}\lambda^{\frac{1}{2}}g^3\mpl$.
\item[(va)] $\phi \gg 13 c_\phi^{-\frac{1}{2}}\lambda^{-1}m$.
\item[(vb)] $\phi \gg 4.4 c_\phi^{-\frac{1}{2}}g^{-2}m$.
\item[(vi)] $m \ll 1.1\times 10^{-4}\gn^{-1}c_\phi^{-1}
   \lambda^{\frac{1}{2}}\mpl.$
\item[(vii)] $\phi \ll 4.8\times 10^{-2}\gn^{-\frac{1}{3}}
   c_\phi^{-\frac{1}{3}}\lambda^{\frac{1}{6}}g^{-1}m^{\frac{2}{3}}
   \mpl^{\frac{1}{3}}$.
\item[(viiia)] $m \gg 1.4\times 10^{-5}\gn^{-1}c_\phi^{-1}\lambda^2\mpl$.
\item[(viiib)] $m \gg 2.7\times
10^{-6}\gn^{-1}c_\phi^{\frac{1}{2}}g^3\mpl.$
\end{description}
Here the conditions (viii) are required to ensure the symmetry
remains broken and warrant the use of the zero-temperature potential,
that is, (viiia) comes from the condition $m^2 \gg \lambda T^2/4$ and
(viiib) from $m^2 \gg c_\phi g^2T^2/12$.

The number of $e$-folds of inflation is expressed as \beq
 N=4.6\times
 10^2\gn^{\frac{1}{3}}c_\phi^{\frac{4}{3}}\lambda^{-\frac{1}{6}}
 m^{-\frac{2}{3}}\mpl^{-\frac{4}{3}}\lmk \phi_f^2 - \phi_i^2 \rmk,
 \label{nib}
\eeq where $\phi_i$ and $\phi_f$ are now the lower and upper bounds
of $\phi$ which satisfy all the above inequalities, respectively.
Inserting (vii) to (\ref{nib}) we find $N \ll
1.1\gn^{-\frac{1}{3}}c_\phi^{\frac{2}{3}}\lambda^{\frac{1}{6}}
g^{-2}m^{\frac{2}{3}}\mpl^{-\frac{2}{3}}$.  The consistency between
(vb) and (vii) sets an upper bound on $m$ as $m \ll 1.3\times
10^{-6}\gn^{-1}c_\phi^{-1}\lambda^{\frac{1}{2}} g^3\mpl$.  These two
inequalities implies $N \ll 1.3\times
10^{-4}\gn^{-1}\lambda^{\frac{1}{2}}$; thus warm inflation is not
feasible.

So far we have used only inequalities (v) and (vii) as promised,
apart from the generic condition (\ref{phiup}) for new inflation. If
we use other inequalities in addition, we can completely close the
allowed region of the parameter space as follows. The consistency
between (\ref{phiup}) and (ii) sets a lower bound on $m$ as \beq
  m \gg 1.1\times
  10^{-8}\gn^{-1}c_\phi^2\lambda^{\frac{7}{2}}\mpl.  \label{mlow}
\eeq
 From (vi) and (\ref{mlow}) we find $\lambda \ll 22c_\phi^{-1}$.
On the other hand, from (\ref{phiup}) and (v) we find $\lambda \gg
1.7\times 10^2c_\phi^{-1}$.  Thus there is no allowed region for
$\lambda$ to realize warm inflation consistently.

Next we move on to  the case viscosity is dominated by Yukawa
interaction, when the temperature is given by \beq
 T=0.20\gn^{-\frac{1}{5}}
 c_\psi^{-\frac{1}{5}}\lambda^{\frac{1}{10}}m^{\frac{2}{5}}
 \phi^{\frac{2}{5}}\mpl^{\frac{1}{5}}.
\eeq The following inequalities must be satisfied.
\begin{description}
\item[(i)]  $\phi \gg 5.5\gn^{\frac{1}{2}}
  c_\psi^{-2}\lambda^{-\frac{3}{2}} m^{4}\mpl^{-3}$.
\item[(ii)] $\phi \ll 2.1\gn^{-\frac{1}{3}}
 c_\psi^{\frac{4}{3}}\lambda^{-\frac{2}{3}}
 m^{\frac{2}{3}}\mpl^{\frac{1}{3}}$.
\item[(iii)] $\phi \ll 2.1\gn^{-\frac{1}{8}}
  c_\psi^{\frac{1}{2}}\lambda^{-\frac{7}{8}}
  m^{\frac{3}{2}}\mpl^{-\frac{1}{2}}$.
\item[(iv)] $\phi \gg 3.0\gn^{-\frac{1}{8}}
  c_\psi^{-\frac{1}{8}}\lambda^{-\frac{1}{4}}
  f^{-\frac{5}{2}}m^{\frac{3}{2}}\mpl^{-\frac{1}{2}}$.
\item[(v)] $\phi \gg 1.4c_\psi^{-\frac{1}{2}}f^{-2}m$.
\item[(vi)] $\phi \gg 56\gn^{\frac{1}{2}}c_\psi^{\frac{1}{2}}
  \lambda^{-\frac{1}{4}} m^{\frac{3}{2}}\mpl^{-\frac{1}{2}}$.
\item[(vii)] $\phi \ll 6.8\times 10^{-2}\gn^{-\frac{1}{3}}
  c_\psi^{-\frac{1}{3}}f^{-\frac{5}{3}}
  \lambda^{\frac{1}{6}}m^{\frac{2}{3}}\mpl^{\frac{1}{3}}$.
\item[(viiia)]$\phi \ll 3.2\times 10^2 \gn^{\frac{1}{2}}
  c_\psi^{\frac{1}{2}}\lambda^{-\frac{3}{2}}m^{\frac{3}{2}}
  \mpl^{-\frac{1}{2}}.$
\item[(viiib)]$\phi \ll 5.3\times 10^2\gn^{\frac{1}{2}}
  c_\psi^{-\frac{3}{4}}f^{-\frac{5}{2}}
  \lambda^{-\frac{1}{4}}m^{\frac{3}{2}}\mpl^{-\frac{1}{2}}.$
\end{description}
 (viiia) comes from the condition $m^2 \gg \lambda T^2/4$ and
(viiib) from $m^2 \gg c_\psi f^2T^2/6$.

In this case, contrary to the case the viscosity arises from bosonic
interactions, there exists some allowed region in the parameter
space, but the number of $e$-folds of inflation, \beq
 N\equiv \int_{\phi_i}^{\phi_f} \frac{H}{\dot{\phi}}d\phi
 =8.0\gn^{-\frac{1}{5}}c_\psi^{\frac{4}{5}}\lambda^{-\frac{2}{5}}
 m^{\frac{2}{5}}\mpl^{-\frac{4}{5}}\lmk \phi_f^{\frac{2}{5}} -
 \phi_i^{\frac{2}{5}}\rmk,
\eeq turns out to be much smaller than unity as seen below. Here
$\phi_i$ and $\phi_f$ are lower and upper bounds on $\phi$ obtained
from the inequalities (i) through (viiib) and (\ref{phiup}) as
before.

 From (vii) we find
\beq
  N \ll 2.7\gn^{-\frac{1}{3}}c_\psi^{\frac{2}{3}}\lambda^{-\frac{1}{3}}
  f^{-\frac{2}{3}}m^{\frac{2}{3}}\mpl^{-\frac{2}{3}}.  \label{nfn}
\eeq Consistency between (v) and (vii) imposes an upper bound on $m$
as $m \ll 1.2\times
10^{-4}\gn^{-1}c_\psi^{\frac{1}{2}}\lambda^{\frac{1}{2}} f\mpl$.
Inserting it to (\ref{nfn}) we find $N \ll 6.6\times
10^{-3}\gn^{-1}c_\psi$.  Thus $N$ is much smaller than unity no
matter how many fermions are interacting with the inflaton.

\subsection{Shifted field model}

So far we have studied the possibilities of chaotic inflation and new
inflation models driven by thermal viscosity term. As a result we
have shown that none of the above models can accommodate inflationary
expansion which lasts more than one $e$-fold.  In fact, we have been
able to rule out them simply from the high-temperature condition
(vii) and the adiabatic condition (v). To cure this problem another
model has been suggested in \cite{bgr} in which the scalar field
$\chi$ has the interaction term \beq
  \CL_{\rm int}=-\sum_j\frac{1}{2}g^2(\phi-M)^2\chi_j^2.
\eeq We are not aware of any particle physics motivation for this
kind of interaction. Nevertheless it may be worthwhile to study this
model because it helps to reduce the effective mass of $\chi$
  when $\phi$ is large and close to $M$. Thus one could hope  that one
may relax a constraint from the high-temperature condition (vii) and
find a sensible solution for warm inflation \cite{bgr}.

The situation, however, is not that simple, as one immediately
recognizes once he writes down the effective equation of motion in
this model.  Indeed instead of eq.\ (\ref{PPPP}), with $C = 96
c_\phi/\pi$, we find
\begin{equation}\label{PPPPPy}
 \ddot{\phi}+\frac{96c_\phi}{\pi T}(\phi-M)^2\dot\phi +
3H\dot{\phi}+m^2\phi+g^2(\phi-M)\sum_j\langle\chi_j^2\rangle_{eq}=0,
\end{equation}
Thus the viscosity term vanishes at $\phi=M$, which makes the
dissipation inefficient in the region $\phi \approx M$.

Nevertheless one may still want to consider the possibility of warm
inflation in the vicinity of $\phi=M$, where the high temperature
condition, $m_\chi=g|\phi-M| \ll T$, is satisfied but the slight
deviation from $\phi=M$ makes the viscosity nonvanishing.  Below we
consider this possibility in chaotic inflation driven by the mass
term for illustration and work out inequalities to be satisfied as we
did above.

In this model the effective potential of $\phi$ is given by \beq
  V_{\rm eff}[\phi] = \frac{1}{2}m^2\phi^2 + \frac{c_\phi
  g^2}{24}T^2(\phi -M)^2+...  \label{smep}
\eeq in the high-temperature limit.  We must treat the cases with
$c_\phi g^2T^2 \ll m^2$ and with $c_\phi g^2T^2 \gg m^2$ separately.

%In particular, the latter case could be important, where the
%thermal mass term plays an important role to determine the
%dynamics of the inflaton even when its contribution to the total
%energy density is sub-dominant.

\subsubsection{ The case $c_\phi g^2T^2 \ll m^2$.}\label{ss}

First we consider the case $c_\phi g^2T^2 \ll m^2$.  Since we are
interested in the regime $\phi > |\phi -M|$, the second term in
(\ref{smep}) is entirely negligible in the inflaton's dynamics. The
effective equation of motion in the slow-roll regime is given by \beq
  C_v\dot\phi=\frac{96c_\phi}{\pi T}(\phi-M)^2\dot\phi=-m^2\phi,
\eeq and the radiation temperature is calculated from
(\ref{radiation}) as \beq
  T=4.3\times 10^{-2}c_\phi^{-\frac{1}{3}}\gn^{-\frac{1}{3}}m
  \phi^{\frac{1}{3}}(\phi-M)^{-\frac{2}{3}}\mpl^{\frac{1}{3}}.
\eeq Apparently it is divergent at $\phi=M$. However, since $\phi=M$
is out of the slow-roll regime and will be automatically excluded
using the inequalities   below, this does not cause any problem.

Now we list the required inequalities relevant to $\chi$:
\begin{description}
\item[(i)] $|\phi-M| \gg 0.17c_\phi^{-\frac{1}{8}}\gn^{-\frac{1}{8}}
    m^{\frac{3}{4}}\phi^{\frac{1}{2}}\mpl^{-\frac{1}{4}}$.
\item[(ii)] $|\phi-M| \gg 8.5\times 10^{-2}c_\phi^{-\frac{1}{2}}
\gn^{-\frac{1}{8}}m^{\frac{3}{4}}\phi^{\frac{1}{8}}\mpl^{\frac{1}{8}}$.
\item[(iii)] $|\phi-M| \gg 5.0\times 10^{-2}\gn^{-\frac{1}{8}}
    m^{\frac{3}{4}}\phi^{-\frac{1}{4}}\mpl^{\frac{1}{2}}$.
\item[(iv)] $|\phi-M| \ll 2.1\times 10^{-7}c_\phi^{-\frac{1}{2}}
    \gn^{-\frac{1}{2}}g^6\phi^{-1}\mpl^2$.
\item[(v)] $|\phi-M| \gg 2.5g^{-2}c_\phi^{-\frac{1}{2}}m$.
\item[(vii)] $|\phi-M| \ll 0.15c_\phi^{-\frac{1}{5}}\gn^{-\frac{1}{5}}
    m^{\frac{3}{5}}\phi^{-\frac{1}{5}}\mpl^{\frac{1}{5}} $.
\end{description}

The number of $e$-folds   of inflation is given by \beq
  N\equiv \int_{|\phi-M|_i}^{|\phi-M|_f}H\frac{d|\phi-M|}{\dot\phi}
 \simeq 4\times 10^2c_\phi^{\frac{1}{3}}\gn^{\frac{1}{3}}m^{-2}
 \mpl^{-\frac{4}{3}}\phi^{-\frac{1}{3}}
 \lmk |\phi-M|_i^{\frac{11}{3}}-|\phi-M|_f^{\frac{11}{3}} \rmk,
 \label{shiftn}
\eeq where the integration has been done near $|\phi-M|$ assuming
$\phi\sim M$.  Inserting (vii) to $|\phi-M|_i$ in (\ref{shiftn}) one
finds \beq
  N \ll 0.4c_\phi^{-\frac{2}{5}}\gn^{-\frac{2}{5}}m^{\frac{1}{5}}
  \phi^{\frac{2}{5}}\mpl^{-\frac{3}{5}}. \label{shiftnn}
\eeq
 From the consistency between (v) and (vii) it follows that
\beq
  m \ll 8.8\times 10^{-4}c_\phi^{\frac{3}{4}}\gn^{-\frac{1}{2}}g^5
  \phi^{\frac{1}{2}}\mpl^{\frac{1}{2}}.
\eeq Using it in (\ref{shiftnn}) one finds that \beq
  N \ll 9\times 10^{-2}c_\phi^{-\frac{1}{4}}\gn^{-\frac{1}{2}}g
  \phi^{\frac{1}{2}}\mpl^{-\frac{1}{2}}.
\eeq This means that unless $\phi\sim M \gg \mpl$ one cannot obtain
sufficiently long period of warm inflation. But then it is not
interesting because for $\phi \gg \mpl$ we can realize chaotic
inflation without the help of thermal viscosity. This conclusion
follows  from (v) and (vii).

But if we consider other inequalities the situation becomes even
worse.
 From the consistency between (i) and (iv) we find
\beq
  m\phi^2 \ll 1.3\times 10^{-8}c_\phi^{-\frac{1}{2}}\gn^{-\frac{1}{2}}g^8
  \mpl^3.
\eeq Inserting it to (\ref{shiftnn}) we obtain \beq
  N \ll 0.01c_\phi^{-\frac{1}{2}}\gn^{-\frac{1}{2}}g^{\frac{8}{5}}.
\eeq Thus warm inflation is ruled out for perturbatively meaningful
values of the coupling constant.

\subsubsection{The case $c_\phi g^2T^2 \gg m^2$.}

Next we consider the opposite limit, $c_\phi g^2T^2 \gg m^2$. If
$g|\phi-M| \gg T$, then all thermal effects disappear. If $g|\phi-M|
\ll T$, one may still neglect contribution of the second term in
(\ref{smep}) to the total energy density of the universe, because
otherwise the universe would be dominated by radiation and there will
be no inflation. This does not mean, however, that this term does not
affect the motion of $\phi$.  On the contrary, the potential force is
dominated by its derivative. As a result the minimum of the potential
is shifted from the origin to the vicinity of $\phi=M$: \beq
  \phi(t)= {M\over 1+{12 m^2\over c_\phi g^2T^2}}
  \cong M-\frac{12m^2M}{c_\phi g^2T^2}. \label{phimin}
\eeq

One may encounter two  possible regimes   by comparing the time scale
of the friction, $\tau_f \equiv C_v^{-1} \propto T(\phi-M)^{-2}$,
with that of $\phi$'s oscillation,
$\tau_o\equiv\sqrt{12}/(c_\phi^{\frac{1}{2}}gT)$.  If the deviation
from $\phi=M$ is sufficiently large to warrant $\tau_f < \tau_o$, the
field is in the slow roll-over regime toward the minimum
(\ref{phimin}) and warm inflation might be possible \footnote{Since
we are interested in the feasibility of viscosity-driven warm
inflation we assume $C_v > 3H$ and hence the only relevant time scale
of friction is $\tau_f = C_v^{-1}$.}. If $\tau_f > \tau_o$, on the
other hand, $\phi$ sits in the minimum (\ref{phimin}) which is
time-dependent through the temperature. Let us consider these two
regimes in turn.

First,   if $\phi$ is rolling toward the minimum, we find the
following effective equation of motion, \beq
 \frac{96c_\phi}{\pi T}(\phi-M)^2\dot\phi=-m^2\phi
 -\frac{c_\phi}{12}g^2T^2(\phi-M).
\eeq
The second term in the right-hand-side (RHS) dominates the
potential force except in the close vicinity of $\phi=M$, and we
neglect the first term in RHS.  In  the regime where the first term
dominates over the second term we would come to  the same conclusion
as in  Sec. \ref{ss}.

 From (\ref{radiation}) one can find  the temperature
\beq
  T=1.8\times 10^6\gn c_\phi^{-1}m\phi \mpl^{-1}.
\eeq We write down the required inequalities relevant to $\chi$.

\begin{description}
\item[(i)] $|\phi-M| \gg 6.0\times 10^2\gn^{\frac{1}{2}}
   c_\phi^{-1}m\phi\mpl^{-1}.$
\item[(ii)] $|\phi-M| \gg 1.6\times 10^{16}\gn^{3}c_\phi^{-3}g^2
   m^2\phi^2\mpl^{-3}.$
\item[(iii)] $\phi \ll 3.1\times
   10^{-14}\gn^{-\frac{5}{2}}c_\phi^2m^{-1}\mpl^2.$
\item[(iv)] $g^4 \gg 3.3\times 10^{-4}c_\phi \gn^{-1}.$
\item[(v)] $|\phi-M| \gg 5.3\times 10^{12}\gn^{2}c_\phi^{-2}m^2
   \phi\mpl^{-2}$.
\item[(vii)] $|\phi-M| \ll 1.8\times 10^6\gn c_\phi^{-1}g^{-1}m\phi
   \mpl^{-1}$.
\end{description}

The number of $e$-folds of inflation is calculated as \beq
  N\equiv\int_{|\phi-M|_i}^{|\phi-M|_f}H\frac{d|\phi-M|}{\dot\phi}
 =6.4\times 10^{-17}\gn^{-3}c_\phi^3m^{-2}\phi^{-2}\mpl
  \lmk |\phi-M|_i^{2}-|\phi-M|_f^{2} \rmk, \label{shiftnc}
\eeq where the integration has been done  over $|\phi-M|$ assuming
$\phi\sim M$. Inserting (vii) to (\ref{shiftnc}) we find $N \ll
2.1\times 10^{-4}\gn^{-1}c_\phi$.  Hence this regime does not lead to
inflation.

Next we suppose that  $\phi$ has fallen to the minimum
(\ref{phimin}).  Since the temperature is presumably gradually
decreasing during the warm inflation, $\phi$ changes with time
according to \beq
  \dot{\phi}\cong\frac{24m^2M}{c_\phi g^2T^3}\dot{T}.
  \label{phidot}
\eeq We would like to see whether this time-variation of $\phi$ may
create a
sufficient amount of radiation by the energy release through the
viscosity term to support quasi-stationary stage of warm inflation.

If such a stage exists at all, we find from (\ref{radiation}) and
(\ref{phidot}) \beq
 \dot{T}=-1.3\times 10^{-2}\gn^{\frac{1}{2}}c_{\phi}^{\frac{3}{2}}
 g^4 m^{-\frac{7}{2}}M^{-\frac{3}{2}}\mpl^{-\frac{1}{2}}T^{\frac{15}{2}}.
\eeq In this case it is convenient to express the required conditions
in terms of the inequalities on the temperature.
\begin{description}
\item[(i)] $T \ll 3.7c_\phi^{-\frac{1}{5}}g^{-\frac{4}{5}}m^{\frac{3}{5}}
   M^{\frac{1}{5}}\mpl^{\frac{1}{5}}.$
\item[(ii)] $T \ll 1.3\gn^{\frac{1}{9}}c_\phi^{\frac{1}{9}}
   g^{\frac{4}{9}}m^{\frac{5}{9}}M^{\frac{1}{3}}\mpl^{\frac{1}{9}}$.
\item[(iii)] $T \ll
   0.32\gn^{-\frac{1}{2}}m^{\frac{1}{2}}M{\frac{1}{2}}$.
\item[(iv)] $T \gg 1.2\times 10^3 g^{-4}mM\mpl^{-1}.$
\item[(v)] $T \ll 0.22\gn^{-\frac{1}{7}}c_\phi^{-\frac{1}{7}}
   g^{\frac{4}{7}}m^{\frac{3}{7}}M^{\frac{3}{7}}\mpl^{\frac{1}{7}}$.
\item[(vii)] $T \gg 2.3 c_\phi^{-\frac{1}{3}}g^{-\frac{1}{3}}
   m^{\frac{2}{3}}M^{\frac{1}{3}}$.
\end{description}
In addition, consistency of the assumption that $\phi$ stays in the
time-dependent minimum (\ref{phimin}) requires $\tau_f > \tau_o$,
namely, \beq
  T > 5.0c_\phi^{-\frac{1}{4}}g^{-\frac{5}{6}}
  m^{\frac{2}{3}}M^{\frac{1}{3}}.  \label{tlow}
\eeq

The $e$-folding number in this regime is expressed by an integral
over the temperature as \beq
  N\equiv \int_{T_i}^{T_f} H\frac{dT}{\dot{T}}
  =25\gn^{-\frac{1}{2}}c_\phi^{-\frac{3}{2}}g^{-4}m^{\frac{9}{2}}
   M^{\frac{5}{2}}\mpl^{-\frac{1}{2}}\lmk T_f^{-\frac{13}{2}}
   - T_i^{-\frac{13}{2}}\rmk, \label{shiftnd}
\eeq where $T_i$ and $T_f$ are the upper and the lower bounds on the
temperature satisfying all the above inequalities, and we have set
$\phi\sim M$. Inserting (\ref{tlow}) to (\ref{shiftnd}) we obtain $N
\ll 7.2\times 10^{-4}\gn^{-\frac{1}{2}}c_\phi^{\frac{1}{8}}
g^{\frac{17}{12}}m^{\frac{1}{6}}M^{\frac{1}{3}}\mpl^{-\frac{1}{2}}$.
Now the consistency between (i) and (iv) imposes an upper bound on
$mM^2$ as $mM^2 \ll 5.3\times 10^{-7}c_\phi^{-\frac{1}{2}}g^8\mpl^3$.
{}From these two inequalities we find $N \ll 6.5\times
10^{-5}\gn^{-\frac{1}{2}}c_\phi^{\frac{1}{24}} g^{\frac{33}{12}}$,
which is much smaller than unity.

Thus we do not see any possibility to implement the idea of warm
inflation even in this model.

\section{Conclusion}\label{conclusion}

In the present paper we have examined feasibility of the warm
inflation scenario \cite{warm} from various view points.

First we discussed how the viscosity term arises in the equation of
motion of a scalar field in a thermal bath following Hosoya and
Sakagami \cite{HS}.  Indeed such a term as $C_v\dot{\phi}$ appears
because it takes  finite time for number density of particles
interacting with
$\phi$
to relax to its thermally equilibrium value while $\phi$ is in motion.

The viscosity term thus obtained could be very large at first glance,
because it is not suppressed by any small coupling constants. As is
evident in its  derivation, however, this term appears as a result of
a small correction to a sub-leading thermal correction term in the
equation of motion.  Since we know that even the leading thermal
correction term does not lead to inflation, such a  sub-leading
viscosity term is not expected to play an important role and to yield
an inflationary regime of a new type.

If one neglects this fundamental feature and solves the overdamped
equation of motion including the term $C_v\dot{\phi}$, one may find
solutions
 indicating   a possible emergence of   warm inflation.  However, we
have found that such solutions  violate   the adiabatic condition
that the scalar field should not change significantly in the
relaxation time of the particles interacting with it. This condition
is necessary for  the derivation of the viscosity term
$C_v\dot{\phi}$. Thus, the equation of motion incorporating the
viscosity term $C_v\dot{\phi}$ fails in the regime where it could
describe warm inflation.

If, on the other hand, we attempt to realize warm inflation in a
manner fully consistent with the field theoretic derivation of the
equation of motion, we  inevitably find that the number of $e$-folds
of inflation is constrained to be much smaller than unity, mainly
 due to the difficulty to satisfy the high-temperature condition and
the adiabatic condition simultaneously.  Even the shifted field
model, which has been proposed to relax the high-temperature
condition \cite{bgr}, turned out to be no exception.

Thus,  our results as a whole show that it is extremely difficult or  
perhaps
even impossible to realize the idea of warm inflation.

\acknowledgements{The authors are thankful to A. Berera for the
discussion  of warm inflation. J.Y. is grateful  to the Institute of
Theoretical
Physics at Stanford University for the hospitality during the time
when this work was done, and to the Monbusho for financial support.
The work by A.L.  was supported in part by NSF grant    PHY-9870115.}

\section{Appendix}

In this appendix we derive the viscosity coefficient arising from
Yukawa interaction. First we calculate the fermion self energy
$\Sigma_\psi$, which is expressed as \beq
  \Sigma_\psi(\vep, \tau_2-\tau_1)=f^2\int\frac{d^3k}{(2\pi)^3}
 S(\tau_2-\tau_1,\vek)G(\tau_2-\tau_1,\vep-\vek), \label{1B}
\eeq in terms of temperature Green function of $\psi$, $S(\tau,\vek)$
and that of $\phi$, $G(\tau,\veq)$.  We use the following spectral
representation of Green functions. \beqa
S(\tau,\vek)&=&\int\frac{d\omega}{2\pi}\sigma(\omega,\vek)e^{-\omega\tau}
 \lkk \lmk
1-n_F(\omega)\rmk\theta(\tau)-n_F(\omega)\theta(-\tau)\rkk\\
 \sigma(\omega,\vek)&=&i\lkk \frac{\omega\gamma_0-\vek\veg+m_\psi}
{(\omega+i\epsilon)^2-\vek^2-m_\psi^2}-\frac{\omega\gamma_0-\vek\veg+m_\psi}
 {(\omega-i\epsilon)^2-\vek^2-m_\psi^2}\rkk \\
  G(\tau,\veq)&=&\int\frac{d\omega}{2\pi}\rho(\omega,\veq)e^{-\omega\tau}
 \lkk -\lmk 1+n_B(\omega)\rmk\theta(\tau)-n_B(\omega)\theta(-\tau)\rkk\\
 \rho(\omega,\veq)&=&
i\lkk \frac{1}{(\omega+i\epsilon)^2-\veq^2-m_\phi^2}
 - \frac{1}{(\omega-i\epsilon)^2-\veq^2-m_\phi^2}\rkk.
\eeqa Inserting them to (\ref{1B}) and applying the Fourier transform
we find the self energy in Matsubara representation \beqa
 \Sigma_\psi(i\omega_l,\vep)&=&\int_0^\beta d\tau e^{i\omega_l\tau}
 \Sigma_\psi(\tau,\vep) \non\\
 &=&-f^2\int\frac{d^3k}{(2\pi)^3}\int\frac{d\omega_1}{2\pi}
 \frac{d\omega_2}{2\pi}\sigma(\omega_1,\vek)\rho(\omega_2,\vep-\vek)
 \frac{1+n_B(\omega_2)-n_F(\omega_1)}{i\omega_l-(\omega_1+\omega_2)},
\eeqa with $\omega_l=(2l+1)\pi T$. The above expression can be
analytically continued to the retarded self energy
$\Sigma_\psi(p_0+i\epsilon,\vep)$, and the imaginary part of the self
energy is given by the discontinuity \beq
  \mbox{Im}\Sigma_\psi(p)=\frac{1}{2i}\lkk \Sigma_\psi(p_0+i\epsilon,\vep)
 - \Sigma_\psi(p_0-i\epsilon,\vep)\rkk
\eeq Explicitly, we find \beqa \mbox{Im}\Sigma_\psi(p)&=&\pi
f^2\int\frac{d^3k}{(2\pi)^3}\int
\frac{d\omega_1}{2\pi}\frac{d\omega_2}{2\pi}\sigma(\omega_1,\vek)
\rho(\omega_2,\vep-\vek)\lmk 1+n_B(\omega_2)-n_F(\omega_1)\rmk
\delta\lmk p_0-\omega_1-\omega_2\rmk \non\\ &=&\pi
f^2\int\frac{d^3k}{(2\pi)^3}\int d\omega_1d\omega_2 \lmk
\omega_1\gamma_0-\vek\veg+m_\psi\rmk \lmk
1+n_B(\omega_2)-n_F(\omega_1)\rmk \mbox{sgn}\omega_1 \delta\lmk
\omega_1^2-\vek^2-m_\psi^2\rmk \non\\ &&\times \mbox{sgn}\omega_2
\delta\lmk \omega_2^2-(\vep-\vek)^2-m_\phi^2\rmk. \eeqa Putting
$p=(p_0,0,0,p_3)$, $|\vek|\equiv k_m$, and $|\vep|\equiv p_m(=p_3)$,
and in the limit $m_\psi$ and $m_\phi$ are negligible compared with
$p_m$, we find \beq
  \mbox{Im}\Sigma_\psi(p)\simeq \frac{\pi}{64}\frac{f^2T^2}{p_m}
  \lmk \gamma_0 - \gamma_3\rmk =\frac{\pi}{64}\frac{f^2T^2}{p_0^2}
  \not{p}\equiv \Ghat\not{p},
\eeq for $p_m=p_0\ltilde T$.

Then the dressed retarded Green function reads \beq
S^R(p_0,\vep)=\frac{1}{(1-i\Ghat)\lmk(p_0+i\epsilon)\gamma^0-\vep\veg\rmk
  -m_{\psi T}},
\eeq in momentum representation, where $m_{\psi T}\equiv m_\psi
+\mbox{Re}\Sigma_\psi$ is the finite-temperature effective mass. The
dressed spectral function is therefore given by \beqa
  \sigma(p)&=& i\lmk S^R(p)-S^{R\dag}(p)\rmk \non\\
 &=& i\lkk \frac{(1-i\Ghat)\not{p}+m_{\psi T}}
 {(1-i\Ghat)^2\lkk (p_0+i\epsilon)^2-\vep^2\rkk-m_{\psi T}^2} -
\frac{(1+i\Ghat)\not{p}+m_{\psi T}}
  {(1+i\Ghat)^2\lkk (p_0-i\epsilon)^2-\vep^2\rkk-m_{\psi T}^2} \rkk\\
 &=& i\lkk \frac{(1-i\Ghat)\not{p}+m_{\psi T}}{2\omega_p(1-i\Ghat)^2}
\lmk\frac{1}{p_0-\omega_p}-\frac{1}{p_0+\omega_p}\rmk-
\frac{(1+i\Ghat)\not{p}+m_{\psi T}}{2\omega_p^\ast (1+i\Ghat)^2}
\lmk\frac{1}{p_0-\omega_p^\ast}-\frac{1}{p_0+\omega_p^\ast}\rmk\rkk,\non
\eeqa with $\omega_p\equiv p_m+im^2_{\psi T}\Ghat/p_m$.  The final
expression applies in the limit $|\vep| \equiv p_m \gg m_{\psi T} .$

In terms of this spectral function the real-time finite-temperature
Green function reads in momentum representation \beqa
&&S^F(t,\vep)=-i\int\frac{dp_0}{2\pi}e^{-ip_0t} \lkk \lmk
1-n_F(p_0)\rmk\theta(t)-n_F(p_0)\theta(-t)\rkk\sigma(p)\\ &=&i\lkk
\frac{(1-i\Ghat)(-\omega_p-\vep\veg)+m_{\psi T}}
{2\omega_p(1-i\Ghat)^2}n_F(\omega_p) e^{i\omega_p t}
+\frac{(1+i\Ghat)(\omega_p^\ast-\vep\veg)+m_{\psi T}}
{2\omega_p(1+i\Ghat)^2}\lmk 1-n_F(\omega_p^\ast)\rmk
e^{-i\omega_p^\ast t}\rkk\theta(t) \non\\ &-&i\lkk
\frac{(1-i\Ghat)(\omega_p-\vep\veg)+m_{\psi T}}
{2\omega_p(1-i\Ghat)^2}n_F(\omega_p) e^{-i\omega_p t}
+\frac{(1+i\Ghat)(-\omega_p^\ast-\vep\veg)+m_{\psi T}}
{2\omega_p^\ast(1+i\Ghat)^2}\lmk 1-n_F(\omega_p^\ast)\rmk
e^{i\omega_p^\ast t}\rkk\theta(-t).\non \eeqa Therefore the reading
term to the viscosity due to Yukawa coupling is given by \beqa
2f^2\dot{\phi}\int^t_{-\infty}dt'(t'-t)\int\frac{d^3p}{(2\pi)^2}
&&\!\!\!\!\mbox{Im}S^F_{\mu\nu}(t-t',\vep)S^F_{\nu\mu}(t'-t,\vep)\non\\
& \simeq &
2f^2\dot{\phi}\int\frac{d^3p}{(2\pi)^2}\int^t_{-\infty}dt'(t'-t)
e^{-2\Gamma(t'-t)}\frac{4m_{\psi T}^2}{p_m^2}\beta\Gamma
\frac{e^{\beta p_m}}{\lmk e^{\beta p_m}+1\rmk^2} \non \\
&=&\frac{64}{\pi^3}\dot{\phi}\int_0^\infty dp_mp_m^3 \frac{e^{\beta
p_m}}{\lmk e^{\beta p_m}+1\rmk^2T^3} \non \\
&=&\frac{288}{\pi^3}\zeta(3)T\dot{\phi}\simeq 11.2T\dot{\phi}, \eeqa
where \beq
  \Gamma_\psi\equiv \frac{m^2_{\psi T}}{p_m}\Ghat=
\frac{\pi f^2 T^2m^2_{\psi T}}{64p_m^3}. \eeq

\end{document}